\documentclass[10pt, conference, letterpaper]{IEEEtran}

\usepackage{array, calc, color, multicol}
\usepackage{algorithm}
\usepackage[noend]{algorithmic} 
\usepackage{graphics, epstopdf, graphicx, epsfig, psfrag, epsf, subfigure}
\usepackage{amssymb, amsfonts, amsmath, times}
\usepackage{multicol,balance, verbatim}
\usepackage{textcomp, mathrsfs, stmaryrd, setspace}
\usepackage{amssymb}

\newcommand{\ie}{{\em i.e., }}

\newcommand{\eg}{{\em e.g., }}

\newtheorem{theorem}{Theorem}

\newtheorem{example}{Example}

\DeclareGraphicsExtensions{.eps, .pdf, .png, .jpg}   

\newcommand{\Nset}{\mathcal{N}}

\newcommand{\Oset}{\mathcal{O}}
\IEEEoverridecommandlockouts

\makeatletter
\newcommand{\oset}[2]{%
{\mathop{#2}\limits^{\vbox to -.5\ex@{\kern-\tw@\ex@
\hbox{\scriptsize #1}\vss}}}}
\makeatother

\begin{document}

\title{Energy-Aware Cooperative Computation in Mobile Devices}

\author{Ajita Singh, Yuxuan Xing, Hulya Seferoglu\\
{ ECE Department, University of Illinois at Chicago}\\
{ \tt asingh64@uic.edu, yxing7@uic.edu, hulya@uic.edu}
}

\maketitle

\begin{abstract}
New data intensive applications, which are continuously emerging in daily routines of mobile devices, significantly increase the demand for data, and pose a challenge for current wireless networks due to scarce resources. Although bandwidth is traditionally considered as the primary scarce resource in wireless networks, the developments in communication theory shifts the focus from bandwidth to other scarce resources including processing power and energy. Especially, in device-to-device networks, where data rates are increasing rapidly, processing power and energy are becoming the primary bottlenecks of the network. Thus, it is crucial to develop new networking mechanisms by taking into account the processing power and energy as bottlenecks. In this paper, we develop an energy-aware cooperative computation framework for mobile devices. In this setup, a group of cooperative mobile devices, within proximity of each other, (i) use their cellular or Wi-Fi (802.11) links as their primary networking interfaces, and (ii) exploit their device-to-device connections (\eg Wi-Fi Direct) to overcome processing power and energy bottlenecks. 
We evaluate our energy-aware cooperative computation framework on a testbed consisting of smartphones and tablets, and we show that it brings significant performance benefits. 
\end{abstract}

\vspace{-5pt}
\section{\label{sec:intro}Introduction}
\vspace{-5pt}
The dramatic increase in mobile applications and the number of devices demanding for wireless connectivity poses a challenge in today's wireless networks \cite{cisco_index, ericsson_report}, and calls for new networking mechanisms. 

\begin{figure}[t!]
\centering
\vspace{-2pt}
\subfigure[The default operation]{ {\includegraphics[height=21mm]{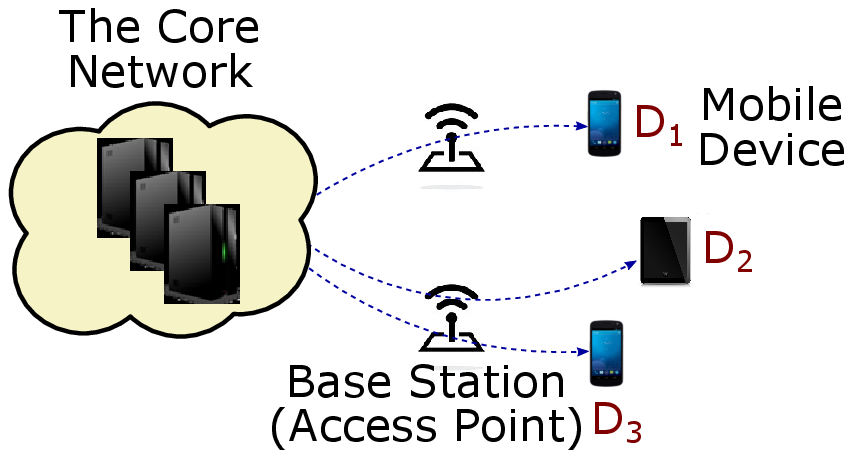}} } 
\subfigure[D2D connectivity]{ {\includegraphics[height=21mm]{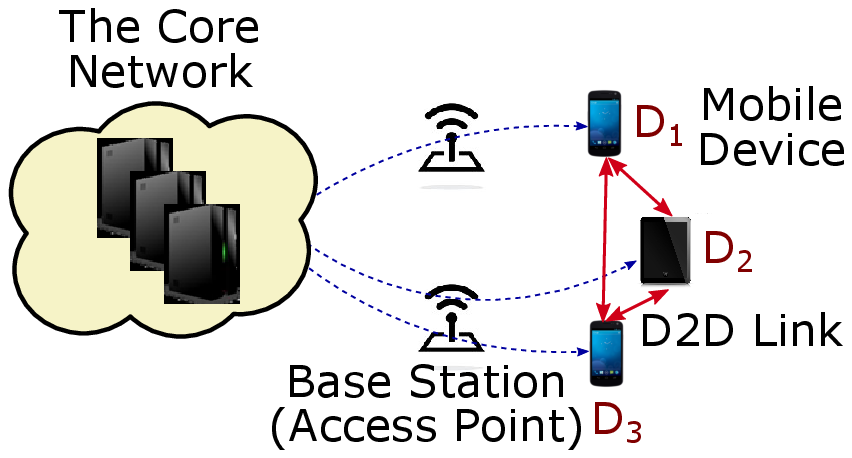}} } 
\vspace{-5pt}
\caption{(a) The default operation for the Internet connection. (b) D2D connectivity: two or more mobile devices can be connected directly, \ie without traversing through the  core network, if they are in close proximity by exploiting local area connections such as Wi-Fi Direct. 
}
\vspace{-18pt}
\label{fig:intro_example_new}
\end{figure}

One of the promising solutions to address the increasing data and connectivity demand is Device-to-Device (D2D) networking. As illustrated in Fig.~\ref{fig:intro_example_new}(a), the default operation in current wireless networks is to  connect each device to the Internet via its cellular or Wi-Fi interface. The D2D connectivity idea, which is illustrated in Fig.~\ref{fig:intro_example_new}(b), breaks this assumption: it advocates that two or more devices in close proximity can be directly connected, \ie without  traversing through auxiliary devices such as a base station or access point. D2D networking, that can be formed by exploiting D2D connections such as Wi-Fi Direct \cite{WiFiDirect}, is a promising solution to the ever increasing number and diversity of applications and devices. In this context, it is crucial to identify scarce resources and effectively utilize them to fully exploit the potential of D2D networking.

Although bandwidth is traditionally considered as the primary scarce resource in wireless networks, in D2D networks, thanks to close proximity among devices and the developments in communication theory, the main bottleneck shifts from bandwidth to other scarce resources including processing power and energy. Next, we present our pilot study demonstrating that processing power can be more pronounced as a bottleneck than bandwidth in D2D networks. 

\begin{figure}[t!]
\centering
\vspace{-3pt}
\subfigure[Setup] { {\includegraphics[width = 35mm]{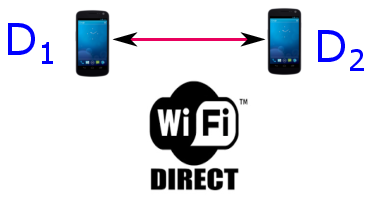}} }  
\subfigure[Rate vs Time]{{\includegraphics[width = 38mm]{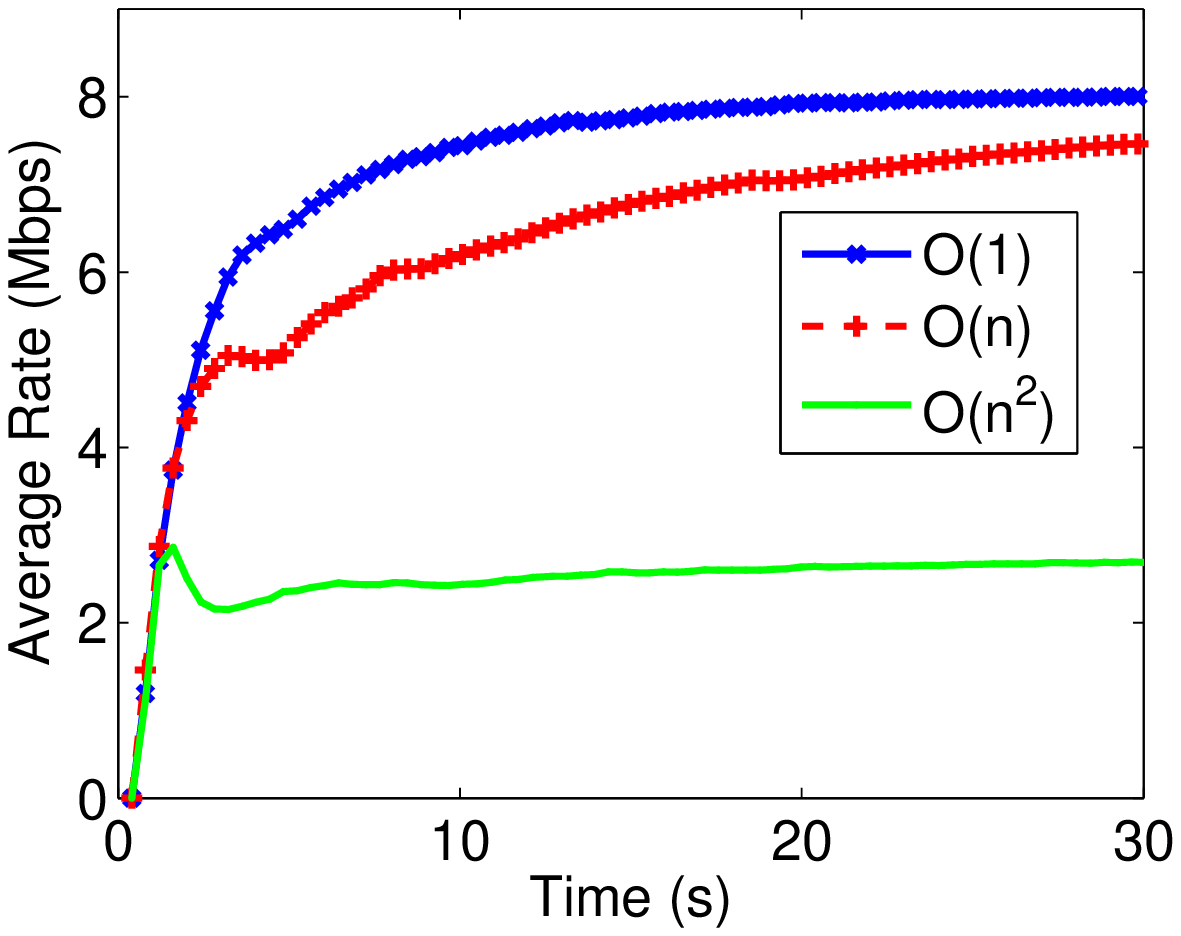}} } 
\vspace{-8pt}
\caption{\textbf{Pilot Study:} (a) Setup: Data is transmitted from mobile device $D_1$ to another mobile device $D_2$. In this setup, 
the mobile devices are Android operating system (OS) \cite{AndroidDeveloper} based Nexus 7 tablets \cite{NexusTechSpecs}. The specific version of the Anroid OS is Android Lollipop 5.1.1. The devices have 16GB storage, 2GB RAM, Qualcomm Snapdragon S4 Pro, 1.5GHz CPU, and Adreno 320, 400MHz GPU. Packet size is $500B$. (b) Transmission rate versus time for different computational complexities at the receiver side. Note that we present the rate that we measure at the mobile device after performing the computations. The presented rates are the average rate of 10 seeds.}
\vspace{-18pt}
\label{fig:complexityVsRate_new}
\end{figure}

\underline{\textit{Pilot Study:}} We developed a prototype for this pilot study as shown in Fig. \ref{fig:complexityVsRate_new}(a), where a mobile device $D_2$ receives data from another device $D_1$ over a Wi-Fi Direct link. We use Android operating system \cite{AndroidDeveloper}  based Nexus 7 tablets \cite{NexusTechSpecs} as mobile devices. In this experiment, after receiving the packets, the mobile device $D_2$ performs operations with complexities of $\Oset(1)$, $\Oset(n)$, and $\Oset(n^2)$ above the transport layer (TCP), where $n$ is the packet size, and the operations we perform are counting the bytes in the packets. In particular, $\Oset(1)$, $\Oset(n)$, and $\Oset(n^2)$ correspond to (i) no counting, (ii) counting every byte in a packet once, and (iii) counting every byte in a packet $n$ times, respectively. We demonstrate in Fig.~\ref{fig:complexityVsRate_new}(c) the received rate at the mobile device $D_2$ (note that this is the rate we measure at the mobile device $D_2$ after performing computations) versus time. This figure demonstrates that the received rate decreases significantly when the complexity increases. \hfill $\Box$

Our pilot study shows that even if actual bandwidth is high and not a bottleneck, processing power could become a bottleneck in D2D networks. Similar observations are made for the energy bottleneck as detailed in Appendix A. Furthermore, with the advances in communication theory, \eg millimeter wave communication \cite{mmWave_rap}, it is expected that  data rates among devices in close proximity will increase significantly, which will make processing power and energy more pronounced as bottlenecks. However, existing applications, algorithms, and protocols are mainly designed by assuming that bandwidth is the main bottleneck. Thus, it is crucial to develop new networking mechanisms when bandwidth is not the primary bottleneck, but processing power and energy are.

Thus, in this paper, our goal is to create group of devices that help each other cooperatively by exploiting high rate D2D connections to overcome the processing power and energy bottlenecks. The next example demonstrates our approach. 
 
\begin{example} \label{ex1} Let us consider Fig.~\ref{fig:intro_example_new}(a) again, where device $D_1$ would like to receive a file from a remote resource via using its cellular or Wi-Fi connection. Assume that the cellular (or Wi-Fi) rates of all devices are 1Mbps, but device $D_1$ can receive data with 500kbps rate due to processing power bottleneck, \ie device $D_1$ has limited processing power (similar to our pilot study we presented earlier). In a traditional system, $D_1$ will behave as a single end point, so its receiving rate will be limited to 500kbps. On the other hand, if devices $D_1$, $D_2$, and $D_3$ will behave as a group and cooperate, then devices $D_2$ and $D_3$ can also receive and process 500kbps portions of data, and transmit the processed data to device $D_1$ over D2D connections. This increases the receiving rate of device $D_1$ to 1.5Mbps from 500kbps, which is a significant improvement. 

This example could be extended for scenarios when energy (battery of mobile devices) is limited.  For example, if device $D_2$'s battery level is too low, its participation to the group activity  should be limited. \hfill $\Box$
\end{example}

\underline{{\em Application Areas.}}  The scenario in the above motivating example could arise in different practical applications from health, education, entertainment, and transportation systems. The following are some example applications. {\em Health:} A person may own a number of health monitoring devices (activity monitoring, hearth monitoring, etc.) which may need updates from the core network. These updates - potentially coded for error correction, compression, and security reasons - should be processed (decoded) by these devices. Processing takes time, which may lead to late reaction to the update (which may require timely response) and energy consumption. On the other hand, by grouping mobile devices, the person's smartphone or tablet could receive the update, process, and pass the processed data to the health monitoring devices via high rate D2D links. {\em Education \& Entertainment:} A group of students may want to watch the video of a lecture from an online education system (or an entertainment video) while sitting together and using several mobile devices. In this setup, one of the devices can download a base layer of a video and decode, while the other devices could download enhancement layers and decode. The decoded video layers could be exchanged among these mobile devices via high rate D2D links. As in the motivating example, if one device's download and decoding rate is limited to 500kbps, it could be improved to 1.5Mbps with the help of other devices. 

Note that the processing overhead in these applications could be due to any computationally intensive task related to data transmission. For example, for video transmission applications, H.264/AVC decoders introduce higher computational complexity when higher quality guarantees are needed \cite{1286980,1218201}. Another example could be network coding; for example, data could be network coded at the source to improve throughput, error correction, packet randomization potential of network coding \cite{NC_meets_TCP}. However, most of the network coding schemes introduce high computational complexity at the receiver side; $\Oset(n^3)$,  \cite{5357971, 5061951}, which limits the transmission rate. Encryption could be another example that introduces processing overhead \cite{modern_complexity}. 

Thus, there exists several applications and scenarios where bandwidth and energy could be bottlenecks, while bandwidth is not the bottleneck. This makes our approach demonstrated in Example~\ref{ex1} crucial. In particular, in this paper, we develop an {\em energy-aware cooperative computation} framework for mobile devices. In this setup, a group of cooperative mobile devices, within proximity of each other, (i) use their cellular or Wi-Fi (802.11) links as their primary networking interfaces, and (ii) exploit their D2D connections (Wi-Fi Direct) for cooperative computation. Our approach is grounded on a network utility maximization (NUM) formulation of the problem and its solution \cite{tutorial_doyle}. The solution decomposes into several parts with an intuitive interpretation, such as flow control, computation control, energy control, and cooperation \& scheduling. Based on the structure of the decomposed solution, we develop a stochastic algorithm; {\em energy-aware cooperative computation}.\footnote{Note that our work focuses on cooperative resource utilization in mobile devices. In this sense, our work is complementary to and synergistic with: (i) creating incentive mechanisms in D2D networks, and (ii) providing privacy and security for D2D users \cite{bits_and_coins, survey_d2d}. Looking into the future, it is very likely that our proposed work on the design, analysis, and implementation of cooperative resource utilization is gracefully combined with the work on creating incentives and providing privacy and security.} 
The following are the key contributions of this work:
\begin{itemize}
  \item We consider a group of cooperative mobile devices within proximity of each other. In this scenario, we first investigate the impact of processing power to transmission rate. Then, we develop an energy-aware cooperative computation  model, where devices depending on their energy constraints could cooperate to get benefit of aggregate processing power in a group of cooperative devices. 
  \item We characterize our problem in a NUM framework by taking into account processing power, energy, and bandwidth constraints. We solve the NUM problem, and use the solution to develop our stochastic algorithm; {\em energy-aware cooperative computation (EaCC)}. We show that EaCC provides stability and optimality guarantees. 
  \item An integral part of our work is to understand the performance of EaCC in practice. Towards this goal, we develop a testbed consisting of  Nexus 5 smartphones and  Nexus 7 tablets. All devices uses Android 5.1.1 as their operation systems. We implement  {\em EaCC} in this testbed, and evaluate it. The experimental results show that our algorithm brings significant performance benefits. 
\end{itemize}

The structure of the rest of the paper is as follows. Section~\ref{sec:related} presents related work. Section~\ref{sec:system} gives an overview of the system model. Section~\ref{sec:NUM} presents the NUM formulation of our cooperative computation scheme. Section~\ref{sec:CoopComp} presents our stochastic algorithm; {\em EaCC}. Section~\ref{sec:performance} evaluates the performance of our scheme in a real testbed. Section~\ref{sec:conclusion} concludes the paper.

\vspace{-5pt}
\section{\label{sec:related} Related Work}
\vspace{-5pt}
This work combines ideas from D2D networking, network utility maximization, and stochastic network control.

The idea of D2D networking is very promising to efficiently utilize resources, so it has found several applications in the literature. In particular, D2D connections are often used to form cooperative groups for data streaming applications, and for the purpose of (i) content dissemination among mobile devices \cite{micro4,micro5}, (ii) cooperative video streaming over mobile devices \cite{microcast, microcast_allerton, micro6, micro7}, and (iii) creating multiple paths and providing better connectivity by using multiple interfaces simultaneously \cite{chesterfield_micro, stiemerling_micro}. As compared to this line of work, we investigate the impact of processing power and energy in D2D networks, and develop mechanisms to effectively utilize these scarce resources. 

D2D networking is often used for the purpose of offloading cellular networks. For example, previous work \cite{micro1, micro2, micro4} disseminate the content to mobile devices by taking advantage of D2D connections to relive the load on cellular networks. Instead of offloading cellular networks, our goal is to create energy-aware cooperation framework to overcome the processing power and energy bottlenecks of mobile devices. 

There is an increasing interest in computing using mobile devices by exploiting connectivity among mobile devices \cite{survey_mobile_to_ubiq}. This approach suggests that if devices in close proximity are capable of processing tasks cooperatively, then these devices could be used together to process a task as it is a cheaper alternative to remote clouds. This approach, sparking a lot of interest, led to some very interesting work in the area \cite{trans_clouds1}, \cite{cloud_down_to_earth}, \cite{mclouds}. As compared to this line of work, we focus on processing power and energy bottlenecks in mobile devices and address the problem by developing energy-aware cooperative computation mechanism. 

An integral part of our proposed work in this task is to develop efficient resource allocation mechanisms. In that sense, our approach is similar to the line of work emerged after the pioneering work in \cite{tass1}, \cite{tass2}, \cite{neelymoli}. However, our focus is on energy-aware cooperative computation.

\vspace{-5pt}
\section{\label{sec:system}System Model}
\begin{figure*}[t!]
\centering
\subfigure[System Model]{ {\includegraphics[height = 34mm]{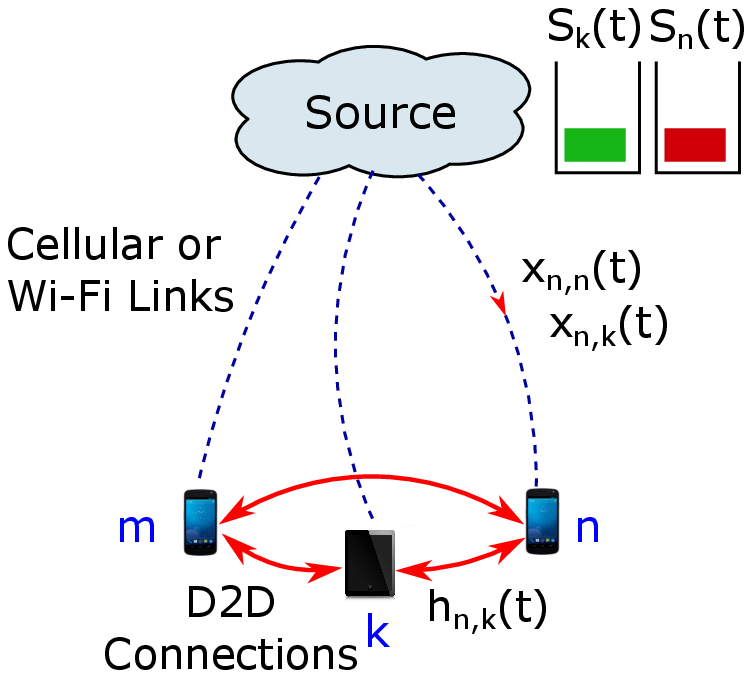}} }   \hspace{2mm}
\subfigure[Building blocks of the source.]{{\includegraphics[height = 25mm]{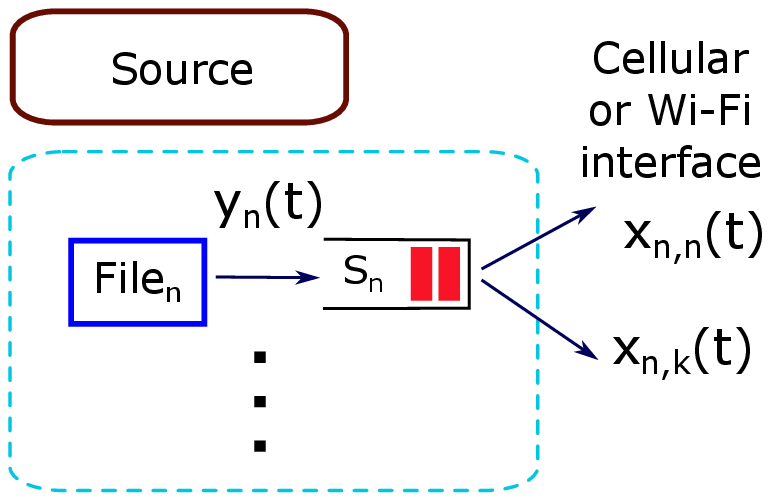}} } \hspace{2mm}
\subfigure[Building blocks of mobile device $n$]{{\includegraphics[height = 33mm]{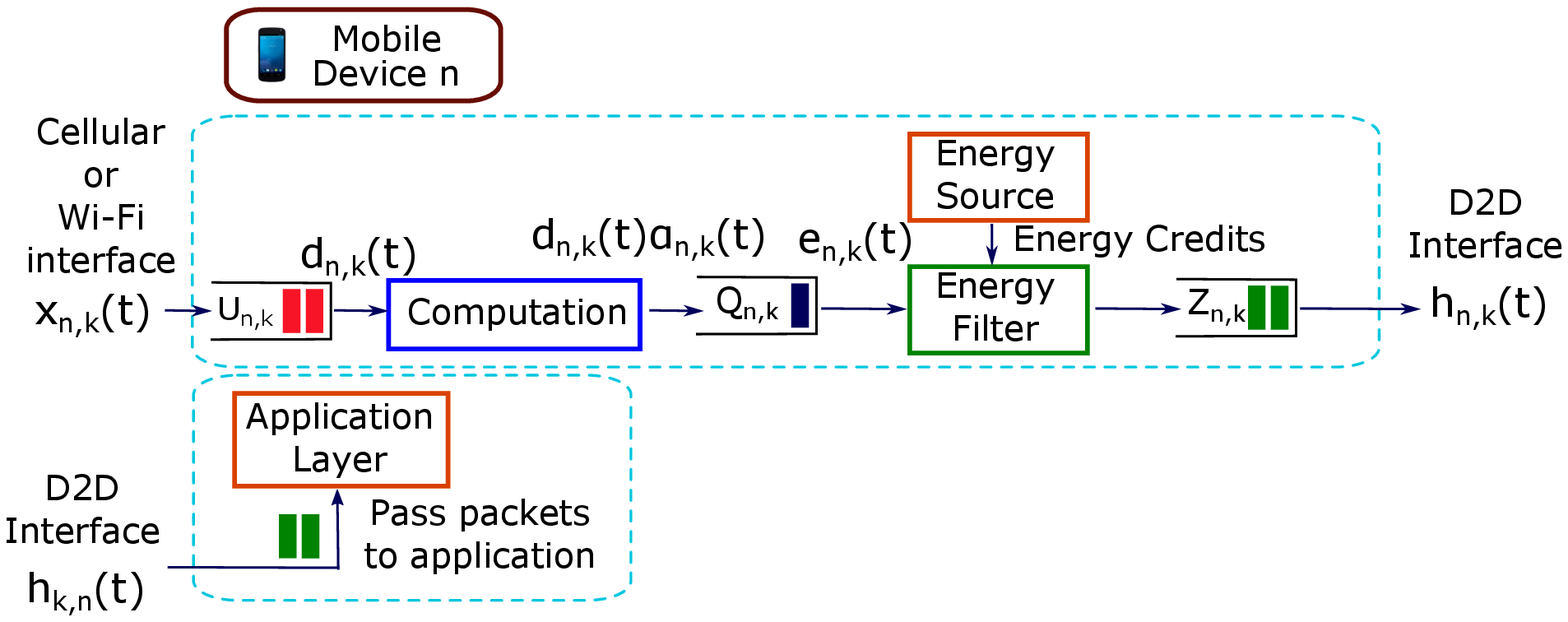}} } 
\vspace{-8pt}
\caption{(a) System model for the scenario of three devices; $n$, $m$, $k$. The source in this model represents the core network and base stations (access points). (b) Building blocks of the source. File$_n$, $\forall n$ is read and inserted in the buffer $S_n(t)$, and packets are transmitted from $S_n(t)$. $x_{n,k}(t)$ is the transmission rate of the packets from the source towards device $n$, and these packets will be processed by device $n$ and forwarded to device $k$.
(c) Building blocks of mobile device $n$. If packets are received from the source via cellular and WiFi interfaces, then they go to the computation and energy control blocks. If packets are received from other mobile devices via D2D interface, they are directly passed to the application. 
}
\vspace{-15pt}
\label{fig:systemModel}
\end{figure*}

\vspace{-5pt}
We consider a cooperative system setup with  $N$ mobile devices, where $\Nset$ is the set of the mobile devices. Our system model for three nodes are illustrated in Fig.~\ref{fig:systemModel}(a). The source in Fig.~\ref{fig:systemModel}(a) represents the core network and base stations (access points). This kind of abstraction helps us focus on the bottlenecks of the system; processing power, energy of mobile devices, and downlink/uplink data rates. In this setup, mobile devices communicate via D2D connections such as Wi-Fi Direct, while the source communicates with mobile devices via cellular or Wi-Fi links. We consider in our analysis that time is slotted and $t$ refers to the beginning of slot $t$. 

{\bf Connecting Devices Together:}
The total flow rate towards device $n$ in  Fig.~\ref{fig:systemModel}(a) (and also explained in Fig.~\ref{fig:systemModel}(b)) is $\sum_{k \in \Nset} x_{n,k}(t) $, where $x_{n,n}(t)$ is the transmission rate of the packets from the source towards device $n$, and these packets will be used by device $n$. Note that $x_{n,k}(t)$ is the transmission rate of the packets from the source towards device $n$, and these packets will be processed by device $n$ and forwarded to device $k$. On the other hand, $y_n(t)$ is the total flow rates targeting device $n$ as demonstrated in Fig.~\ref{fig:systemModel}(b). The source constructs a queue $S_{n}(t)$ for the packets that will be transmitted to the mobile device $n$. The evolution of $S_n(t)$ based on $y_n(t)$ and $x_{k,n}(t)$ is expressed as 
\begin{align} \label{eq:queueS}
S_{n}(t+1) \leq \max [S_n(t) - \sum_{k \in \Nset} x_{k,n}(t), 0] + y_{n}(t), 
\end{align} where the inequality comes from the fact that there may be less than $y_{n}(t)$ packets arriving into $S_{n}(t)$ at time $t$ in practice (\eg in real time applications, the number of available packets for transmission could be limited). 

The flow rate $y_n(t)$ is coupled with a utility function $g_n(y_n(t))$, which we assume to be strictly concave function of $y_n(t)$. This requirement is necessary to ensure stability and utility optimality of our algorithms. The ultimate goal in our resource allocation problem is to determine the flow rates; $y_n(t)$ which maximize the sum utility $\sum_{n \in \Nset} g_n(y_n(t))$.  

Finally, flow rate over D2D connection between device $n$ and $k$ is $h_{n,k}(t)$, $k\neq n$. Note that $h_{n,k}(t)$ is to help node $k$ using node $n$ as a processing device.

{\bf Inside a Mobile Device:} In each device, we develop different modules depending on where data is arriving from (as shown in Fig.~\ref{fig:systemModel}(c)); \ie from the source via cellular or WiFi interface, or other mobile devices via D2D interfaces. 

When data is arriving from a D2D interface, it is directly passed to the application layer, as this data is already processed by another device. On the other hand, when data is arriving from the source via cellular or WiFi interfaces, packets go through multiple queues as shown in Fig.~\ref{fig:systemModel}(c), where $U_{n,k}$, $Q_{n,k}$, and $Z_{n,k}$ represent three different queues constructed at mobile device $n$ for the purpose of helping node $k$. Incoming packets via cellular or WiFi links are stored in $U_{n,k}$, which then forwards the packets to {\em computation} block with rate $d_{n,k}(t)$. The computation block processes the packets, and pass them to queue $Q_{n,k}$. Note that the output rate from computation block is $d_{n,k}(t) \alpha_{n,k}(t)$, where $\alpha_{n,k}(t)$ is a positive real value. This value captures any possible rate changes at the computation block, \ie $\alpha_{n,k}(t)$ is a rate shaper. For example, if the computation block is H.264/AVC decoder or transcoder, we expect that the rate at the output of the computation block should be higher than the input. Thus, $\alpha_{n,k}(t)$ captures this fact for any $n,k,t$. On the other hand, if there is no rate change after the processing, then $\alpha_{n,k}(t) = 1$.

The processed (and possibly rate shaped) packets are queued at $Q_{n,k}(t)$ and passed to {\em energy filter}. The energy filter is coupled to the energy source, which determines the amount of energy that can be spent to support the tasks at each slot. The amount of energy is determined according to {\em  energy credits}. In particular, the energy source, depending on the battery level as well as the estimate on the expected battery consumption in the near future, calculates the number of packets that can be supported by the mobile device, and the same number of energy credits enter the energy filter. (Note that both energy filter, energy source, and energy credits are not real, but virtual entities, so they can be modeled by using a few counters in practice.) Thus, at each transmission slot, packets are transmitted from $Q_{n,k}(t)$ to $Z_{n,k}(t)$ if there are energy credits in the energy filter. Finally, packets from $Z_{n,k}(t)$ are transmitted to application if device $n$ is the destination of the data (\ie $n=k$), or they are transmitted to the original destination via D2D interface with rate $h_{n,k}(t)$. 

The computation and energy filter blocks in Fig.~\ref{fig:systemModel}(c) model the processing and energy bottlenecks of the mobile device, respectively. If packets in $U_{n,k}$ increases too much, this means that the computation block, hence processing power, is the bottleneck, so node $n$ should not receive much packets from the source. Similarly, if $Q_{n,k}$ increases too much, this means that energy filter is the bottleneck, so again node $n$ should not receive much packets. Note that there could be also some buildup in $Z_{n,k}$ if the link between node $n$ and $k$ is the bottleneck of the system, and it should be taken into account when the energy-aware cooperative computation framework is developed. 

Also, it is crucial in our system model to put energy filter after the computation block, because if device $n$ will help device $k$, the actual amount of packets that are supposed to be transmitted are the processed packets, which will cause energy consumption (\ie not the packets before processing). 

Based on the above intuitions and observations, we will develop our resource allocation problem and algorithm in the next sections. The evolution of the queues $U_{n,k}(t)$, $Q_{n,k}(t)$, and $Z_{n,k}(t)$ are provided in Table~\ref{table:queueEvolution}. 

\begin{table}[t!]
\vspace{-5pt}
\caption{Evolution of queues $U_{n,k}(t)$, $Q_{n,k}(t)$, and $Z_{n,k}(t)$.}
\centering
\begin{tabular}{  | c | }
    \hline  $U_{n,k}(t+1) \leq \max [U_{n,k}(t) - d_{n,k}(t), 0] + x_{n,k}(t)$  \\
    \hline  $Q_{n,k}(t+1) \leq \max [Q_{n,k}(t) - e_{n,k}(t), 0] + d_{n,k}(t) \alpha_{n,k}(t)$    \\
    \hline  $Z_{n,k}(t+1) \leq \max [Z_{n,k}(t) - h_{n,k}(t), 0] + e_{n,k}(t)$   \\
    \hline
\end{tabular}
\label{table:queueEvolution}
\vspace{-5pt}
\end{table}

{\bf Links:}
In our system model, we consider two scenarios: (i) cellular + Wi-Fi Direct, and (ii) Wi-Fi + Wi-Fi Direct.  In both cases, the D2D links between mobile devices are Wi-Direct. In the first case, \ie in cellular + Wi-Fi Direct, the links between the source and mobile devices are cellular, while they are Wi-Fi in the second case, \ie in  Wi-Fi + Wi-Fi Direct. These two scenarios are different from each other, because in the first scenario, cellular and Wi-Fi Direct links could operate simultaneously as they use different parts of the spectrum. On the other hand, in the second scenario, both Wi-Fi and Wi-Fi Direct use the same spectrum, so they time share the available resources. Our model and energy-aware cooperative computation framework are designed to operate in both scenarios. Next, we provide details about our link models.\footnote{Note that the link models described in this section provide a guideline in our algorithm development and basis in our theoretical analysis. However, in Section~\ref{sec:performance}, we relax the link model assumptions we made in this section, and evaluate our algorithms on real devices and using real links.}
 
In the system model in Fig.~\ref{fig:systemModel}(a), each mobile device $n \in \Nset$ is connected to the Internet via its cellular or Wi-Fi link. At slot $t$, $\boldsymbol C^{s}(t)$ is the channel state vector of these links, where $\boldsymbol C^{s}(t) = \{C_{1}^{s}(t),$ $...,$ $C_{n}^{s}(t),$ $...,$ $C_{N}^{s}(t)\}$. We assume that $C_{n}^{s}(t)$ is the state of the link between the source and mobile device $n$, and it takes ``ON'' and ``OFF'' values depending on the state of the channel. Without loss of generality, if mobile device $n$ does not have Internet connection, then $C_{n}^{s}(t)$ is always at ``OFF'' state, which means there is no cellular or Wi-Fi connection. 

Since we consider that mobile devices are in close proximity and transmission range, they form a fully connected clique topology. At slot $t$, $\boldsymbol C^{w}(t)$ is the channel state vector of the D2D links, where $\boldsymbol C^{w}(t)$ $=$ $\{C_{1,2}^{w}(t),$ $...,$ $C_{n,k}^{w}(t),$ $...,$ $C_{N-1,N}^{w}(t)\}$. We assume that $C_{n,k}^{w}(t)$ is the state of the D2D link between node $n$ and $k$. 

We consider protocol model in our formulations \cite{gupta_interference_model}, where each mobile device can either transmit or receive at the same time at the same frequency. Assuming that $\boldsymbol C(t) = \{\boldsymbol C^{s}(t), \boldsymbol C^{w}(t)\}$ is the channel state vector of the system including both the links between the source and mobile devices as well as among mobile devices, $\Gamma_{\boldsymbol C(t)}$ denotes the set of the link transmission rates feasible at time slot $t$ depending on our protocol model. In particular, for cellular + Wi-Fi Direct setup, $\Gamma_{\boldsymbol C(t)}$ is the set that allows more links to be operated at the same time, while for the Wi-Fi + Wi-Fi Direct setup, $\Gamma_{\boldsymbol C(t)}$ is more limited set to take into account the interference among the links.

\vspace{-5pt}
\section{\label{sec:NUM} Problem Formulation}
\vspace{-5pt}
In this section, we characterize the stability region of the energy-aware cooperative computation problem, and formulate network utility maximization (NUM) framework. The solution of the NUM framework provides us insights for developing the stochastic control algorithms in the next section.\footnote{Note that NUM optimizes the average values of the parameters that are defined in Section~\ref{sec:system}. By abuse of notation, we use a variable, \eg $\phi$ as the average value of $\phi(t)$  in our NUM formulation if both $\phi$ and $\phi(t)$ refers to the same parameter.}

\subsection{\label{sec:stabRegion} Stability Region}
We provide the stability region of the cooperative computation system for both cellular + Wi-Fi Direct and W-Fi + Wi-Fi Direct setups. First, the flow conservation constraint at the source should be $y_n \leq \sum_{k \in \Nset} x_{k,n}$ to stabilize the system. This constraint requires that the total outgoing rate from the source, \ie $\sum_{k \in \Nset} x_{k,n}$ should be larger than the generated rate $y_n$.  

Furthermore, the following flow conservation constraints inside a mobile device should be satisfied for stability; $x_{n,k} \leq d_{n,k}$, $d_{n,k} \alpha_{n,k} \leq e_{n,k}$, and $e_{n,k} \leq h_{n,k}$. These constraints are necessary for the stability of queues $U_{n,k}$,  $Q_{n,k}$, and  $Z_{n,k}$, respectively. Finally, the transmission rates over the links should be feasible, \ie $\{ x_{n,k}, h_{n,k} \}_{\forall n \in \Nset, k \in \Nset} \in \Gamma_{\boldsymbol C}$. 

Thus, we define the stability region as $\Lambda = \{ \{y_n, x_{n,k}, d_{n,k}, e_{n,k}, h_{n,k}\}_{\forall n \in \Nset, k \in \Nset}$ $|$ $y_n$, $x_{n,k}$, $d_{n,k}$, $e_{n,k}$, $h_{n,k}$ $\geq 0$, $\forall n \in \Nset, k \in \Nset$, $y_n \leq \sum_{k \in \Nset} x_{k,n}$, $x_{n,k} \leq d_{n,k}$, $d_{n,k} \alpha_{n,k} \leq e_{n,k}$, $e_{n,k} \leq h_{n,k}$, $\{ x_{n,k}, h_{n,k} \}_{\forall n \in \Nset, k \in \Nset} \in \Gamma_{\boldsymbol C}\}$.

\subsection{\label{sec:NUM_Formulation} NUM Formulation}
Now, we characterize our NUM problem. 
\begin{align} \label{opt:eq1}
\max_{\boldsymbol y} \mbox{ } & \sum_{n \in \Nset} g_n(y_n)  \nonumber \\
\mbox{s.t.} \mbox{ }  & y_n, x_{n,k}, d_{n,k}, e_{n,k}, h_{n,k} \in \Lambda,  \mbox{ } \forall n \in \Nset, k \in \Nset
\end{align}
The objective of the NUM problem in (\ref{opt:eq1}) is to determine  $y_n, x_{n,k}, d_{n,k}, e_{n,k}, h_{n,k}$ for $\forall n \in \Nset, k \in \Nset$ which maximize the total utility $\sum_{n \in \Nset} g_n(y_n)$. 

\subsection{\label{sec:NUM_Solution} NUM Solution}
Lagrangian relaxation of the flow conservation constraints that characterize the stability region $\Lambda$ gives the following Lagrange function:

\begin{align} \label{relax:eq1}
L & = \sum_{n \in \Nset} g_{n}(y_{n}) - \sum_{n \in \Nset} s_n (y_n - \sum_{k \in \Nset} x_{k,n}) - \sum_{n \in \Nset} \sum_{k \in \Nset} u_{n,k} \nonumber \\
&  (x_{n,k} - d_{n,k}) - \sum_{n \in \Nset} \sum_{k \in \Nset} q_{n,k}(d_{n,k}\alpha_{n,k} - e_{n,k}) - \sum_{n \in \Nset} \sum_{k \in \Nset} \nonumber \\
& z_{n,k} (e_{n,k} - h_{n,k}) 
\end{align} where $s_n$, $u_{n,k}$, $q_{n,k}$, and $z_{n,k}$ are the Lagrange multipliers. Note that we will convert these Lagrange multipliers to queues $S_n$, $U_{n,k}$, $Q_{n,k}$, and $Z_{n,k}$ when we design our stochastic algorithm in the next section.

The Lagrange function in (\ref{relax:eq1}) is decomposed into sub-problems such as flow, computation, and energy controls as well as cooperation and scheduling. The solutions of  (\ref{relax:eq1}) for $y_n$, $d_{n,k}$, $e_{n,k}$, $x_{n,k}$, and $h_{n,k}$ are expressed as: 
\begin{align} 
\label{eq:decSol1} & \mbox{\underline {Flow control:} } \max_{\boldsymbol y} \mbox{ }   \sum_{n \in \Nset} (g_{n}(y_n) - y_n s_n)  \\
\label{eq:decSol2} & \mbox{\underline {Computation control:} }\max_{\boldsymbol d} \mbox{ }  \sum_{n \in \Nset} \sum_{k \in \Nset} d_{n,k} (u_{n,k} - q_{n,k} \alpha_{n,k})   \\
\label{eq:decSol3}  & \mbox{\underline {Energy control:} } \max_{\boldsymbol e} \mbox{ }  \sum_{n \in \Nset} \sum_{k \in \Nset} e_{n,k} (q_{n,k} - z_{n,k})  \\
\label{eq:decSol4} & \mbox{\underline {Cooperation \& Scheduling:} } \nonumber \\
& \max_{\boldsymbol x, h} \mbox{ }  \sum_{n \in \Nset} \sum_{k \in \Nset} [x_{n,k} (s_k - u_{n,k}) + z_{n,k} h_{n,k}] \nonumber \\
& \mbox{s.t. } \mbox{ } \{ x_{n,k}, h_{n,k} \}_{\forall n \in \Nset, k \in \Nset} \in \Gamma_{\boldsymbol C}
\end{align} 

Next, we design a stochastic algorithm; energy-aware cooperative computation inspired by the NUM solutions in (\ref{eq:decSol1}), (\ref{eq:decSol2}), (\ref{eq:decSol3}), (\ref{eq:decSol4}).

\section{\label{sec:CoopComp} Energy-Aware Cooperative Computation }
Now, we provide our energy-aware cooperative computation algorithm  which includes {\em flow control}, {\em computation control}, {\em energy control}, and  {\em cooperation \& scheduling}. 

\underline{Energy-Aware Cooperative Computation (EaCC):}
\begin{itemize}
 \item {\em Flow Control:} At every time slot $t$, ${y}_n(t)$ is determined by maximizing $\max_{\boldsymbol {{y}}}$  $[M g_{n}({y}_n(t)) -  {S}_{n}(t) {y}_{n}(t) ]$ subject to  ${y}_n(t) \leq R_{n}^{\max}$, where $R_{n}^{\max}$ is a positive constant larger than the transmission rate from the source, and $M$ is a large positive constant. Note that ${S}_{n}(t)$ is the queue size at the source of flow and stores packets that are supposed to be transmitted to mobile device $n$. After ${y}_n(t)$ is determined, ${y}_n(t)$ packets are inserted in queue $S_n(t)$ (as illustrated in Fig.~\ref{fig:systemModel}(a)). 

\item {\em Computation Control:}  At every time slot $t$, the computation control algorithm at device $n$ determines $d_{n,k}(t)$ by optimizing
\begin{align} \label{eq:decoder}
\max_{\boldsymbol d} & \mbox{  } \sum_{k \in \Nset} d_{n,k}(t) [U_{n,k}(t) - Q_{n,k}(t)\alpha_{n,k}(t)] \nonumber \\
\mbox{s.t. } & \sum_{k \in \Nset} d_{n,k}(t) \leq D_{n}^{\max}
\end{align} where $D_{n}^{\max}$ is a positive constant larger than the processing rate of the computation block in device $n$ dedicated to help device $k$. The interpretation of (\ref{eq:decoder}) is that at every time slot $t$, $d_{n,k^{*}}= D_{n}^{\max}$ packets are passed to the computation block (in Fig.~\ref{fig:systemModel}(b)) if $U_{n,k^{*}}(t) - Q_{n,k^{*}}(t) > 0$, where $k^{*}$ is the mobile device that maximizes (\ref{eq:decoder}).  Otherwise, no packets are sent to the computation block. The packets that are being processed by the computation block are passed to $Q_{n,k}(t)$. Note that some computation blocks may require to receive a group of packets to be able to process them. In that case, $D_{n}^{\max}$ is arranged accordingly (\ie it can be increased to transfer a group of packets).

\item {\em Energy Control:} 
At every time slot $t$, the energy control algorithm at device $n$ determines $e_{n,k}(t)$ by optimizing 
\begin{align} \label{eq:energy}
\max_{\boldsymbol e} & \mbox{  } \sum_{k \in \Nset} e_{n,k}(t) [Q_{n,k}(t) - Z_{n,k}(t)] \nonumber \\
\mbox{s.t. } & \sum_{k \in \Nset} e_{n,k}(t) \leq E_{n,k}^{\max}
\end{align} where $E_{n}^{\max}$ is a positive constant larger than the energy capacity of device $n$ dedicated to help device $k$. The interpretation of (\ref{eq:energy}) is that at every time slot $t$, $e_{n,k^{*}}= E_{n}^{\max}$ packets are passed to the energy filter (as illustrated in Fig.~\ref{fig:systemModel}(b)) if $Q_{n,k^{*}}(t) - Z_{n,k^{*}}(t) > 0$, where $k^{*}$ is the mobile device that maximizes (\ref{eq:energy}). Otherwise, no packets are sent to the energy filter. The packets passing through the energy filter are inserted in $Z_{n,k}(t)$.
    
\item {\em Scheduling \& Cooperation:}  At every time slot $t$, the scheduling and cooperation algorithm determines transmission rates over links, \ie $x_{n,k}(t)$ and $h_{n,k}(t)$ by maximizing 
\begin{align} \label{eq:schedulingCoop}
\max_{\boldsymbol x, \boldsymbol h} & \mbox{  } \sum_{n \in \Nset} \sum_{k \in \Nset} [x_{n,k}(t) (S_{k}(t) - U_{n,k}(t)) + h_{n,k}(t) Z_{n,k}(t) ] \nonumber \\
\mbox{s.t. } & {\boldsymbol x, \boldsymbol h} \in \Gamma_{\boldsymbol C(t)}
\end{align} For cellular + Wi-Fi Direct system, (\ref{eq:schedulingCoop}) is decomposed into two terms: maximizing $\sum_{n \in \Nset} \sum_{k \in \Nset}$ $x_{n,k}(t) (S_{k}(t)$ $-$ $U_{n,k}(t))$ and $\sum_{n \in \Nset} \sum_{k \in \Nset}$ $h_{n,k}(t) Z_{n,k}(t)$, because cellular and Wi-Fi Direct transmissions operate simultaneously and transmission over one link does not affect the other. On the other hand, for Wi-Fi + Wi-Fi Direct setup, the joint optimization in (\ref{eq:schedulingCoop}) should be solved. 

Note that transmissions over all links are unicast transmissions in our work, where unicast is dominantly used in practice over cellular, Wi-Fi, and Wi-Fi Direct links. Also, it is straightforward to extend our framework for broadcast transmissions. 
 
\end{itemize}

\begin{theorem}\label{eec_theorem1}
If channel states are i.i.d. over time slots, and the arrival rates $E[y_{n}(t)] = A_n, \forall n \in \Nset$ are interior of the stability region $\Lambda$, then energy-aware cooperative computation stabilizes the network and the total average queue sizes are bounded.

Furthermore, if the channel states are i.i.d. over time slots, and the traffic arrival rates are controlled by the flow control algorithm of energy-aware cooperative computation, then the admitted flow rates converge to the utility optimal operating point with increasing $M$.
\end{theorem}
{\em Proof:} The proof is provided in Appendix B. $\blacksquare$

Our energy-aware cooperative computation framework has several advantages: (i) distributed, (ii) takes into account scarce resources such as processing power and energy in addition to bandwidth to make control decisions, and (iii) utilizes available resources; processing power, energy, and bandwidth in a utility optimal manner. Theorem~\ref{eec_theorem1} shows the theoretical performance guarantees of our framework, while we focus on its performance in a practical setup in the next section. 

\vspace{-5pt}
\section{\label{sec:performance} Performance Evaluation}
\vspace{-5pt}
In this section, we evaluate our {\em energy-aware cooperative computation} (EaCC) scheme using a tested that consists of Android based smartphones and tablets. The evaluation results show that our scheme significantly improves throughput as compared to (i) no-cooperation, where each device receives its content from the source without cooperating with other devices,  and (ii) cooperation, where multiple mobile devices cooperate, but the cooperating devices do not do computation and energy control for other devices (mobile devices just receive packets from the source, and relay them to other mobile devices without processing and energy control). Next, we present testbed setup and results in detail. 

\begin{figure*}[t!]
\centering
\vspace{-5pt}
\subfigure[System Model]{ {\includegraphics[height = 32mm]{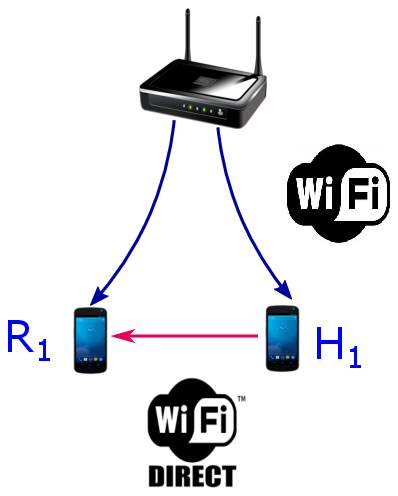}} } 
\subfigure[Rate vs. Time]{ {\includegraphics[height = 32mm]{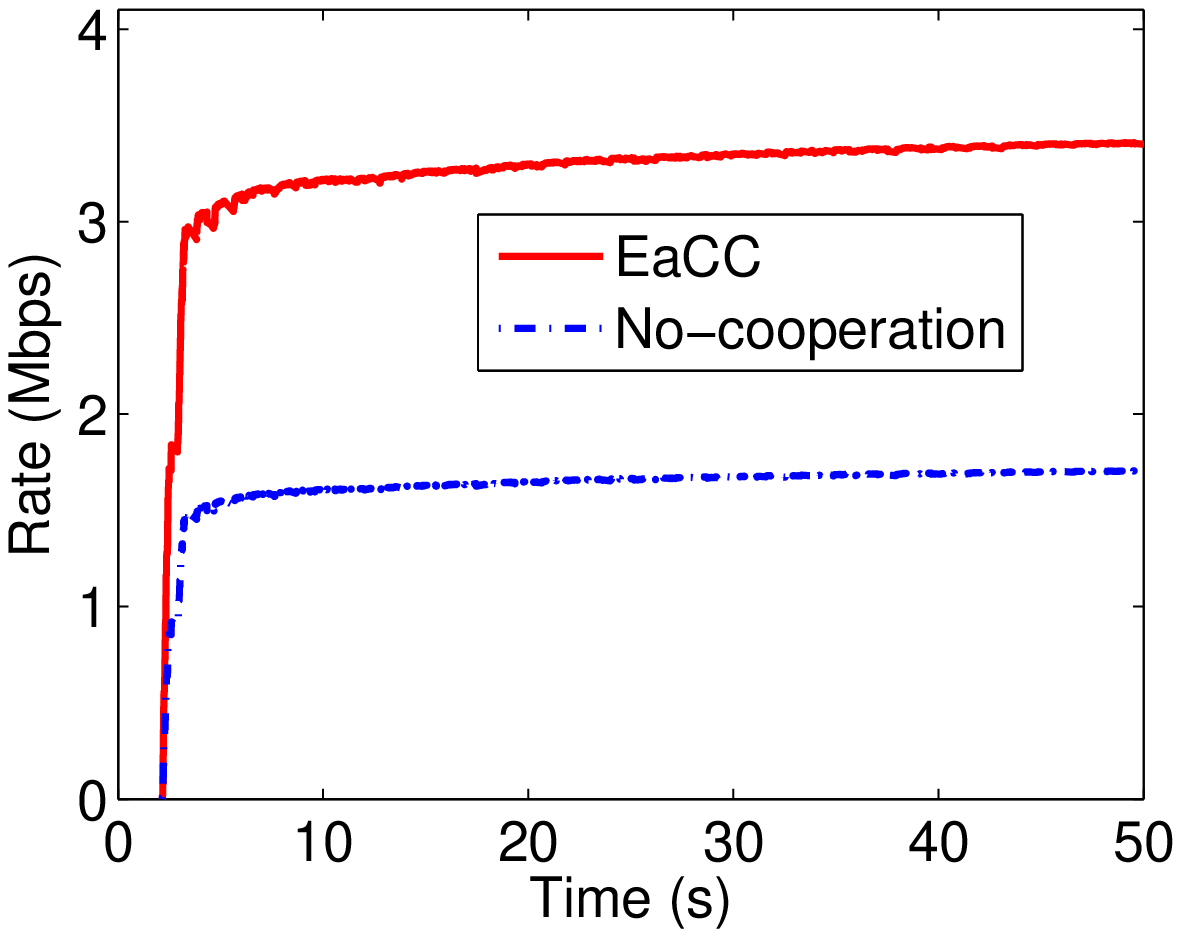}} } 
\subfigure[Rate vs. Energy]{ {\includegraphics[height = 32mm]{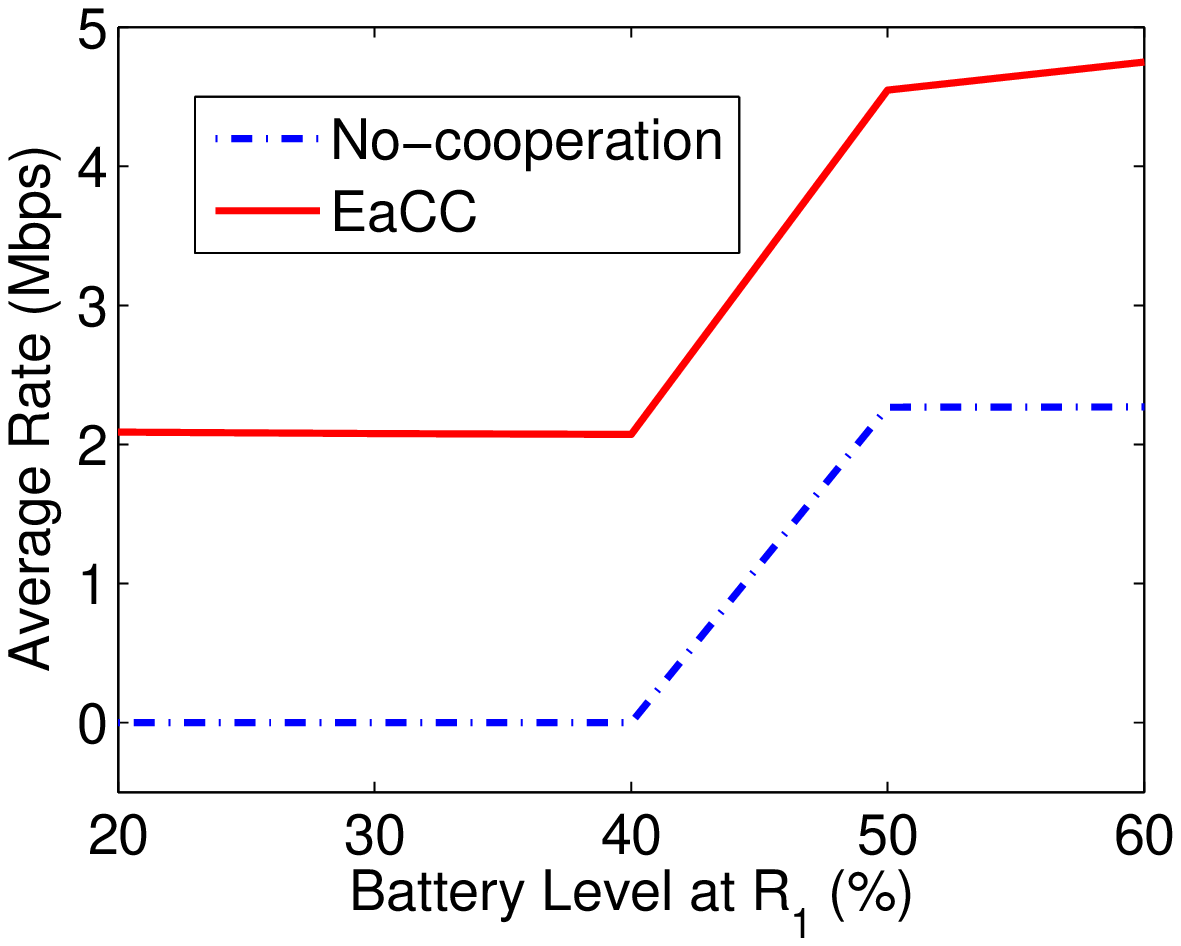}} } 
\vspace{-5pt}
\caption{(a) System model consisting of a source device, one receiver, and one helper. Wi-Fi is used between the source and the receiver device ($R_1$) and the helper device ($H_1$), while Wi-Fi Direct is used to connect $R_1$ to $H_1$. (b) Average rate versus time for the setup shown in (a) for the case that all the devices are Android  based Nexus 7 tablets. (c) Average rate versus energy level at receiver device $R_1$. In this setup, all devices are Android based Nexus 5 smartphones. In both (b) and (c), the average rate is calculated as the average over 10 trials (with different seeds). The computation under consideration in this experiment is $O(n^2)$, which basically counts the number of bytes in a packet for each byte in the packet (\ie recursive counting).
}
\vspace{-15pt}
\label{fig:Rate_vs_time_results}
\end{figure*}

\subsection{Setup \& Implementation Details}
\vspace{-5pt}
{\em Devices:} We implemented a testbed of the setup shown in Fig.~\ref{fig:systemModel}(a) using real mobile devices, specifically Android 5.1.1 based Nexus 5 smartphones and Nexus 7 tablets. 

We classify devices as (i) a source device, which acts as the source in Fig.~\ref{fig:systemModel}(a), (ii) helper devices, which receive data from the source, process it, and transmit to other devices (receivers) to help them, and (iii) receiver devices, which receive data from both the source device and the helpers. A receiver device processes data arriving from the source, but it does not process the data arriving from helpers as the helpers send already processed data. Note that a device could be both receiver and a helper device depending on the configuration. 
 
{\em Integration to the Protocol Stack:} We implemented our energy-aware cooperative computation (EaCC) framework as a slim layer between transport and application layers. In other words, we implemented our framework on top of TCP. This kind of implementation is advantageous, because (i) mobile devices do not need rooting, and (ii) our framework and codes could be easily transferred to mobile devices using other operating systems such as iOS.\footnote{We note that we will make our codes and applications publicly available, which we believe will contribute to the research community.} 

{\em Source Configuration and EaCC Implementation:} We implemented the source node in Fig.~\ref{fig:systemModel} using a Nexus 5 smartphone. 
Basically, multiple files; File$_n$, File$_k$ requested by devices $n$ and $k$ are read by using the public java class {\em BufferedInputStream} according to the flow control algorithm described in Section~\ref{sec:CoopComp} and shown in Fig.~\ref{fig:systemModel}(b). The bytestream is packetized by setting each packet to $500B$, and packets are inserted into source buffers; $S_n(t)$, $S_k(t)$.  We set the flow control parameters as; $M=500$, $R_n^{\max} = 100$, and slot duration is $20msec$. We used $\log$ function as our utility function. In this setup, reading files, converting bytestream into packets, and inserting packets into the input queues are done by multiple threads, \ie a thread runs for each file; File$_n$ in Fig.~\ref{fig:systemModel}(b).

The other set of threads at the source device make packet transmission decisions from the source device to receiver and helper devices. In particular, the source node collects $U_{n,k}(t)$ information from all mobile devices.  At each time slot, the source node checks $S_k(t) - U_{n,k}(t)$, and if $S_k(t) - U_{n,k}(t) > 0$, then $100$ packets are transmitted from $S_k(t)$ to the TCP socket at the source device for transmission to mobile device $n$.

{\em EaCC Operation on Mobile Devices:} All mobile devices (including helper or receiver+helper devices) implement all the building blocks illustrated Fig.~\ref{fig:systemModel}(c). Multiple threads are used to make these blocks operating simultaneously.

The first thread at mobile device $n$ receives packets that are transmitted by the source node, and inserts these packets in $U_{n,k}$. 

The second thread has two tasks. First, it transfers packets from $U_{n,k}$ to $Q_{n,k}$ according to the computation control algorithm in (\ref{eq:decoder}), where $D_{n}^{\max} = 100$ packets and the slot duration is $20msec$. We set $\alpha_{n,k}(t)=1$ in our experiments as our applications do not change the rate as explained later in this section.  The second task of this thread to actually do the computation tasks related to the application. In our experiments, the computation block counts the bytes in the packets. In particular, similar to the pilot study in the introduction, $\Oset(1)$, $\Oset(n)$, and $\Oset(n^2)$ correspond to (i) no counting, (ii) counting every byte in a packet once, and (iii) counting every byte in a packet $n$ times, respectively. 

The third thread transfers packets from $Q_{n,k}$ to $Z_{n,k}$ using the energy control algorithm in (\ref{eq:energy}), where we set $E_{n,k}^{\max}$ depending on the battery level of the device. For example, if the battery level is below some threshold $E_{n,k}^{\max}$ is limited. We evaluated different configurations in our experiments as we explain layer. The slot duration is again set to $20msec$. 

The final thread transfers packets from $Z_{n,k}$ to application layer if $n=k$, or transmits to node $k$ if $n \neq k$. In the second case, \ie if $n \neq k$, the number of packets in TCP socket is checked at every time slot, where the time slot duration is $20msec$. If it is below a threshold of $500$ packets, then $100$ packets are removed from $Z_{n,k}$ and inserted to the TCP socket to be transmitted to node $k$. 

When node $n$ receives packets from node $k$, it directly passes the packets to the application layer as illustrated in Fig.~\ref{fig:systemModel}(c), because these packets are the ones that are already processed by node $k$. If node $n$ is both a helper and a receiver device, it runs all the threads explained above in addition to the receiving thread from node $k$ (illustrated in Fig.~\ref{fig:systemModel}(c)). 

{\em Information Exchange:} Our implementation is lightweight in the sense that it limits control information exchange among mobile devices. The only control information that is transmitted in the system is $U_{n,k}$ from each mobile device to the source node. Each mobile device $n$ collects $U_{n,k}$, $\forall k \in \Nset$, and transmits this information to the source node periodically, where we set the periods $100msec$. 

{\em Connections:} All the devices in the system including the source device, helpers, receivers, and helper+receiver devices are connected to each other using Wi-Fi Direct connections in our testbed. The source node is configured as the group owner of the Wi-Fi Direct group. We note that cooperation in this setup does not bring any benefit in terms of bandwidth utilization as all the links use the same transmission channel in a Wi-Fi Direct group. However, as we demonstrate later in this section, it brings benefit due to cooperative processing power and energy utilization, which is our main focus in this paper. Therefore, this setup (where all the devices are connected to each other using Wi-Fi Direct links) well suits to our evaluation purposes.

{\em Test Environment:} We conducted our experiments using our testbed in a lab environment where several other Wi-Fi networks were operating in the  background. We located all the devices in close proximity of each other, and we have evaluated EaCC for varying levels of computational complexity, number of receivers, and number of helpers. Next, we present our evaluation results. 

\begin{figure*}[t!]
\centering
\vspace{-5pt}
\subfigure[System Model]{ {\includegraphics[height = 32mm]{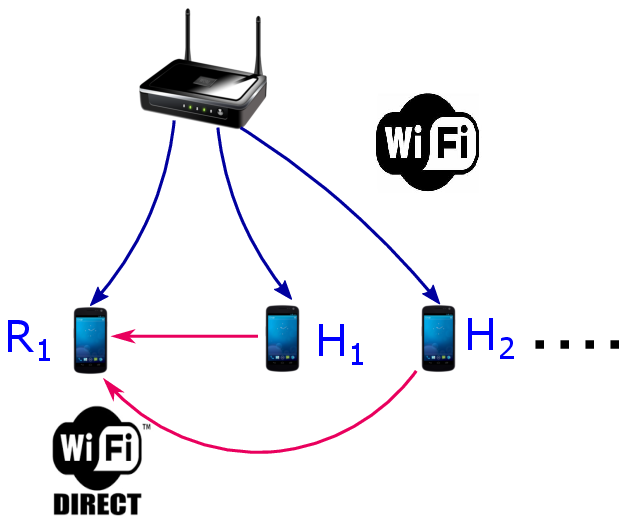}} } 
\subfigure[Rate vs. Number of Helpers]{ {\includegraphics[height = 32mm]{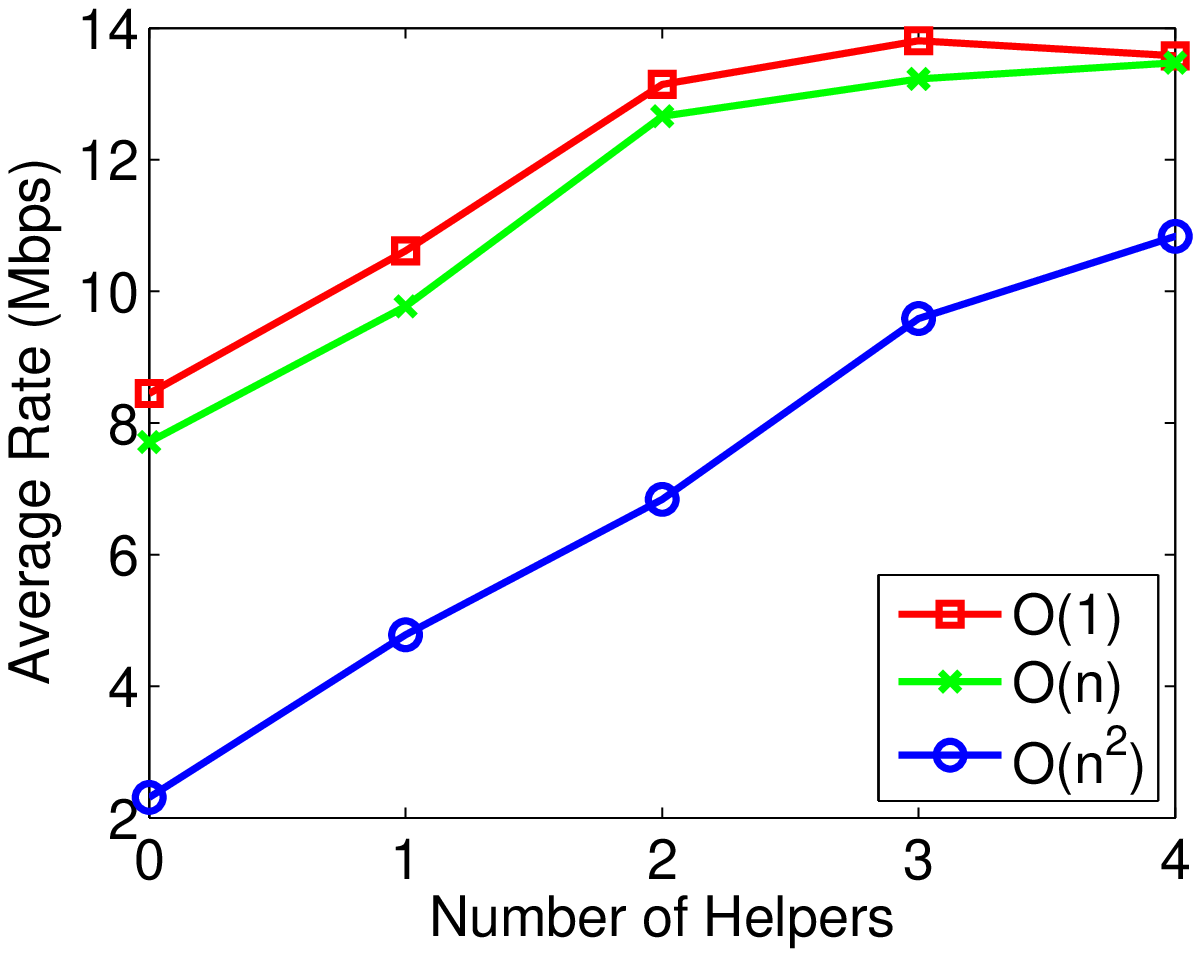}} } 
\subfigure[Rate vs. Number of Helpers]{ {\includegraphics[height = 32mm]{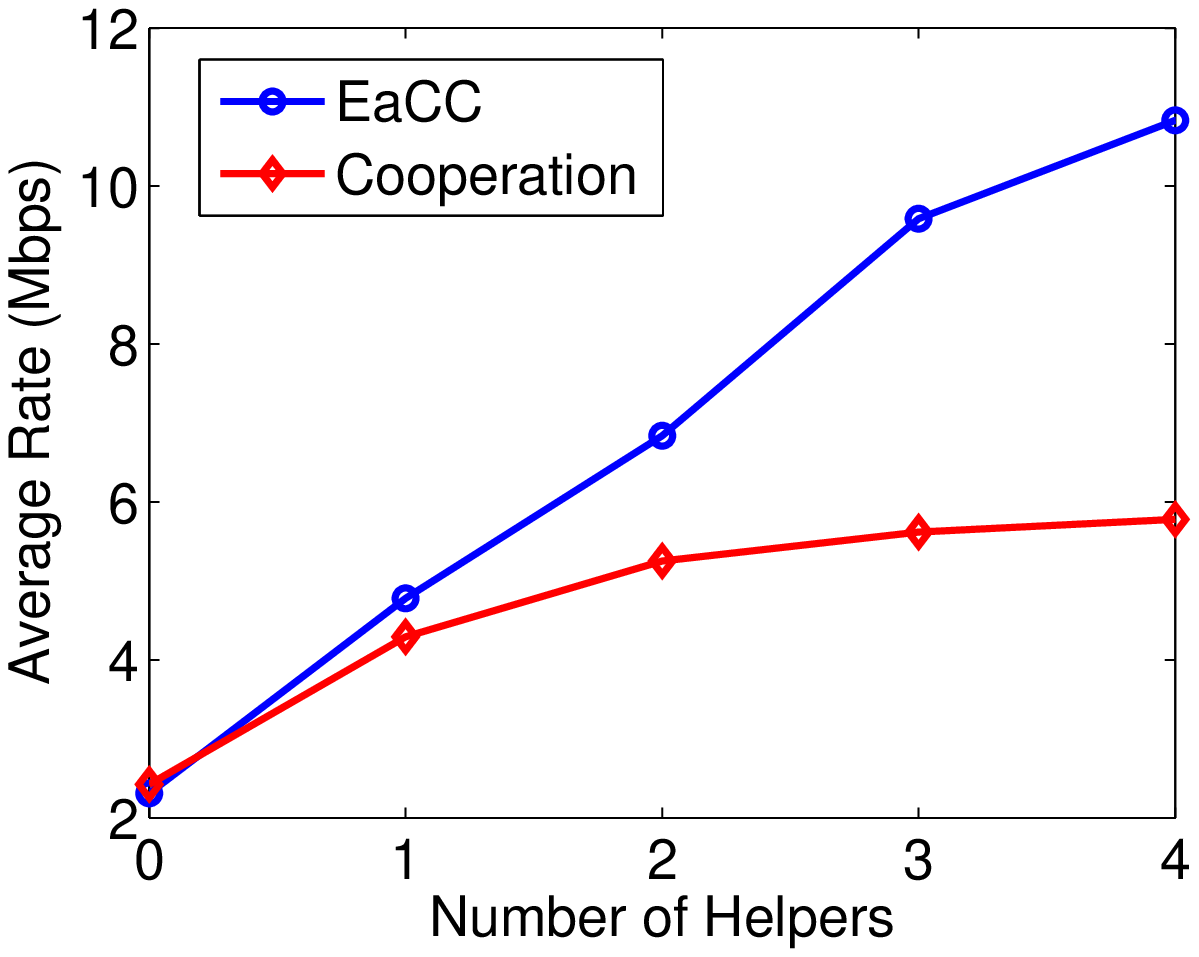}} } 
\vspace{-5pt}
\caption{(a) System model consisting of a source device, one receiver, and multiple helpers. Wi-Fi is used between the source and the receiver device ($R_1$) and the helper devices ($H_1, \ldots$), while Wi-Fi Direct is used to connect the receiver devices with the helper devices. (b) EaCC: Average rate measured at receiver $R_1$ versus the number of helpers. (c) Average rate measured at receiver $R_1$ versus the number of helpers for EaCC and cooperation, when the complexity is $O(n^2)$. 
}
\vspace{-15pt}
\label{fig:Rate_vs_helpers_one_receiver_results}
\end{figure*}

\subsection{Results}
\vspace{-5pt}
We first consider a setup as shown in Fig.~\ref{fig:Rate_vs_time_results}(a) which consists of a source device, one receiver ($R_1$), and one helper ($H_1$). Fig.~\ref{fig:Rate_vs_time_results}(b) shows the average rate versus time graph for the setup shown in Fig.~\ref{fig:Rate_vs_time_results}(a) when all three devices are Android based Nexus 7 tablets. The average rate is calculated as the average over 10 trials (with different seeds). The computation under consideration in this experiment is $O(n^2)$, which basically counts the number of bytes in a packet $n$ times, $n$ is the packet size. As can be seen, if there is no cooperation, the rate measured at $R_1$ is on the order of 1.5Mbps. On the other hand, EaCC increases the rate to almost 3Mbps. This means that helper device $H_1$ helps the receiver device $R_1$ to process the packets in EaCC. In this setup, EaCC doubles the rate as compared to no-cooperation, which is a significant improvement. 

For the same setup in Fig.~\ref{fig:Rate_vs_time_results}(a), we also evaluate the effect of energy control part of EaCC on the average rate performance. In particular, Fig.~\ref{fig:Rate_vs_time_results}(c) shows the average rate versus battery level at the receiver device $R_1$. In these results, we used Android based Nexus 5 smartphones. The average rate is calculated as the average over 10 trials (with different seeds). The computation under consideration in this experiment is $O(n^2)$, which basically counts the number of bytes in a packet for each byte in the packet. We consider that if the battery level of a device reduces below 40\% threshold, then energy credits are not generated for the processing of the received packets. This makes $Q_{n,k}$ large over time, and after some point no packets are transmitted to that device for the processing task. In Fig.~\ref{fig:Rate_vs_time_results}(c), when the battery level of $R_1$ reduces below 40\%, then it stops receiving packets for processing. If there is no cooperation, then the rate towards $R_1$ reduces to 0. On the other hand, with EaCC, the rate is still higher than 0 thanks to having helper. The helper device with larger energy level (for the sake of this experiment), receives packets from the source, process them, and forwards them to $R_1$, which receives already processed data. After 40\% threshold, both EaCC and no-cooperation improve, because $R_1$ starts processing packets. This result shows the importance of energy-awareness in our cooperative computation setup.

Now, we consider the effect of the number of helpers to overall rate performance. In particular, we develop a setup shown in Fig.~\ref{fig:Rate_vs_helpers_one_receiver_results}(a), where there is one source, one receiver, and varying number of helpers. In this setup, the source device, receiver, and the first two helper devices are Nexus 5 smartphones, while the other helpers are Nexus 7 tablets. Fig.~\ref{fig:Rate_vs_helpers_one_receiver_results}(b) shows the average rate (averaged over 10 seeds) when EaCC is employed versus the number of helpers for different computational complexities such as $O(1)$, $O(n)$, and $O(n^2)$, where the processing task is counting the number of bytes in a packet. As expected, when complexity increases, the rate decreases. More interestingly, the increasing number of helpers increases the rates of all complexity levels. There are two reasons for this behavior. First, even if complexity level is low, \eg $O(1)$, processing power is still a bottleneck, and it can be solved by increasing the number of helpers. Note that after the number of helpers exceeds a value, the achievable rates saturate, which means that processing power is not a bottleneck anymore, but bandwidth is. The second reason is that receiving data over multiple interfaces increases diversity. In other words, when the channel condition over one interface (\eg between source and the mobile device) degrades, the other interface (\eg between two mobile devices) can still have a better channel condition. 

In order to understand the real impact of processing power in a cooperative system, we tested both EaCC and cooperation (without computation and energy control) in the setup shown in Fig.~\ref{fig:Rate_vs_helpers_one_receiver_results}(a). The results are provided in Fig.~\ref{fig:Rate_vs_helpers_one_receiver_results}(c) when the complexity is $O(n^2)$. As can be seen, while EaCC significantly increases the rate with increasing number of helpers, cooperation slightly increases the rate (due to diversity). The improvement of EaCC over cooperation is as high as 83\%, which is significant.

\begin{figure*}[t!]
\centering
\vspace{-5pt}
\subfigure[System Model]{ {\includegraphics[height = 32mm]{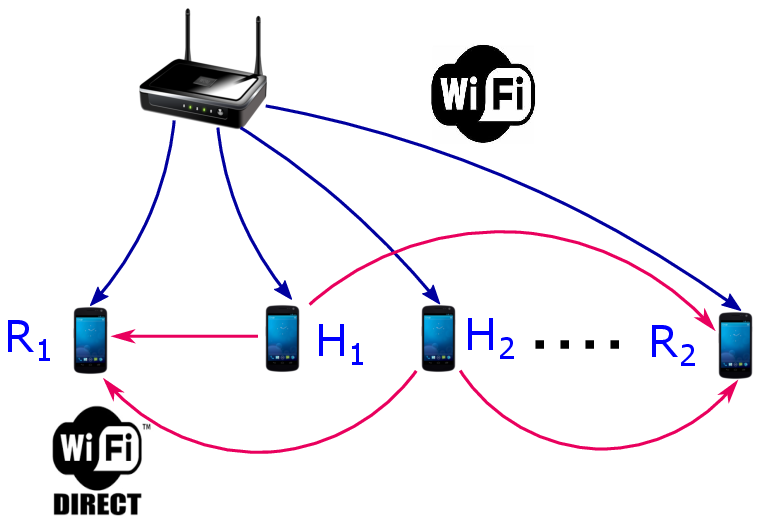}} } 
\subfigure[Rate at $R_1$ vs. Number of Helpers]{ {\includegraphics[height = 32mm]{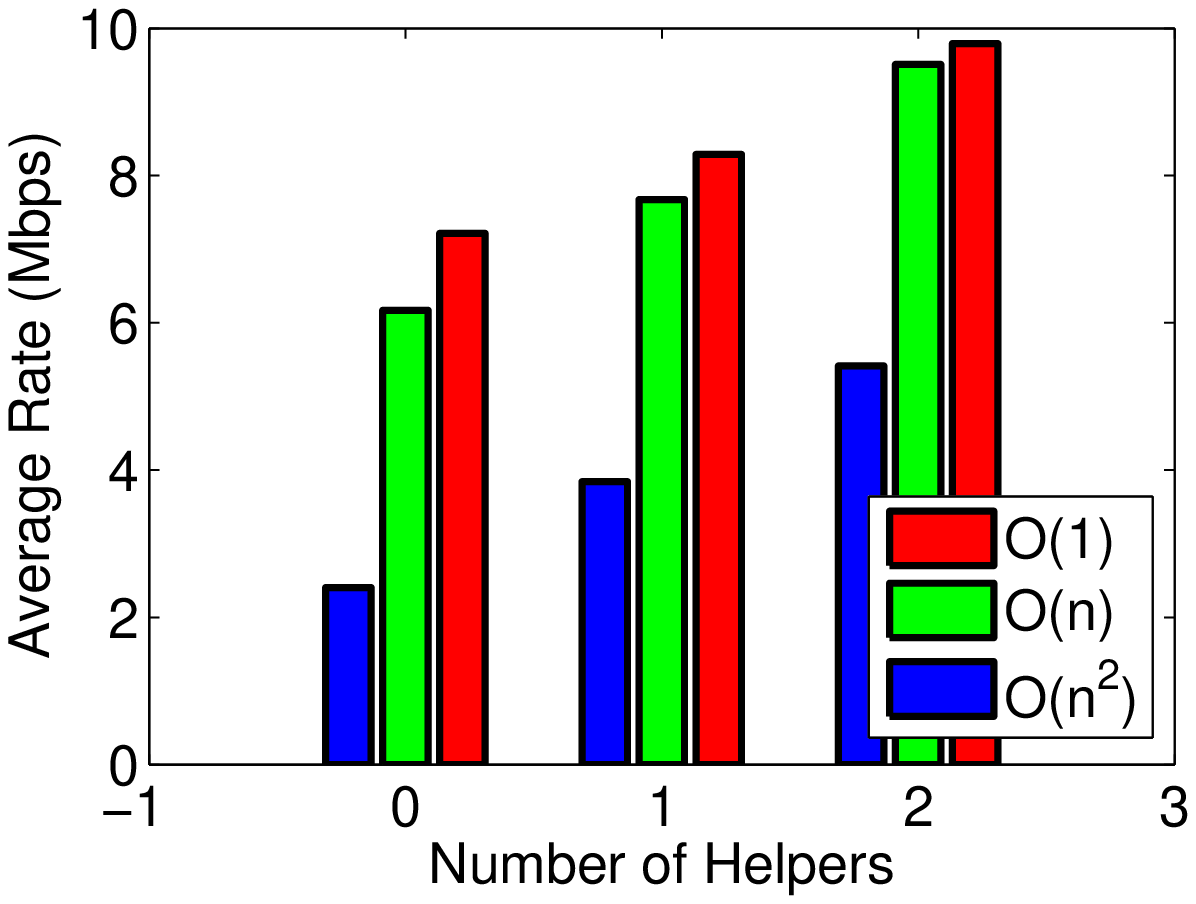}} } 
\subfigure[Rate at $R_2$ vs. Number of Helpers]{ {\includegraphics[height = 32mm]{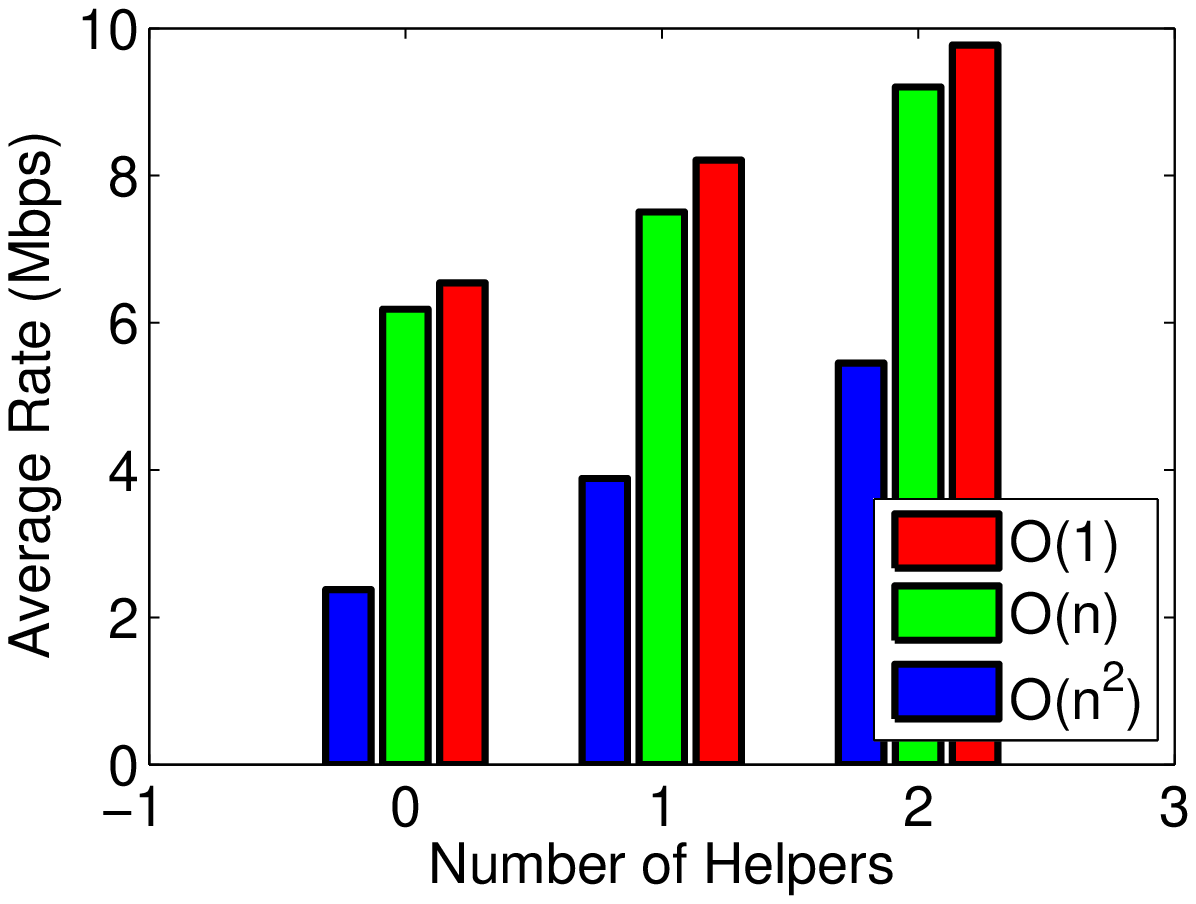}} } 
\vspace{-5pt}
\caption{(a) System model consisting of a source device, two receivers, and multiple helpers. Wi-Fi is used between the source and the receiver devices ($R_1$, $R_2$) and the helper devices ($H_1, \ldots$), while Wi-Fi Direct is used to connect the receiver devices with the helper devices. (b) EaCC: Average rate measured at receiver $R_1$ versus the number of helpers. (c)  EaCC: Average rate measured at receiver $R_2$ versus the number of helpers.}
\vspace{-15pt}
\label{fig:Rate_vs_helpers_two_receiver_results}
\end{figure*}

Finally, we consider a scenario that there are multiple receivers interested in different files. Fig.~\ref{fig:Rate_vs_helpers_two_receiver_results}(a) shows the system model with one source, two receivers, and multiple helpers. In this setup, the source, two receivers, and the first helper is Android based Nexus 5 smartphone, while the rest of the helpers are Nexus 7 tablets. Fig.~\ref{fig:Rate_vs_helpers_two_receiver_results}(b) and (c) show the average rate (averaged over 10 seeds) measured at $R_1$ and $R_2$ when EaCC is employed with respect to the increasing number of helpers, respectively. Similar to previous setups, $O(1)$, $O(n)$, and $O(n^2)$ correspond to different computational complexities, where the processing task is counting the number of bytes in a packet. As can be seen, the measured rate at both $R_1$ and $R_2$ increases with increasing number of helpers. This shows that our EaCC algorithm successfully accommodates multiple flows and receivers.

\vspace{-5pt}
\section{\label{sec:conclusion}Conclusion}
\vspace{-5pt}
We considered that a group of cooperative mobile devices, within proximity of each other, (i) use their cellular or Wi-Fi (802.11) links as their primary networking interfaces, and (ii) exploit their D2D connections (Wi-Fi Direct) for cooperative computation. We showed that if mobile devices cooperate to utilize their aggregate processing power, it significantly improves transmission rates. Thus, for this scenario, we developed an {\em energy-aware cooperative computation} framework to effectively utilize processing power and energy. This framework provides a set of algorithms including flow, computation and energy controls as well as cooperation and scheduling. We implemented these algorithms in a testbed which consists of real mobile devices. The experiments in the testbed show that our {\em energy-aware cooperative computation} framework brings significant performance benefits.

\bibliographystyle{IEEEtran}


\section*{Appendix A: Energy Bottleneck}

\begin{figure*}[t!]
\centering
\vspace{-5pt}
\subfigure[$\Oset(1)$]{ {\includegraphics[height = 32mm]{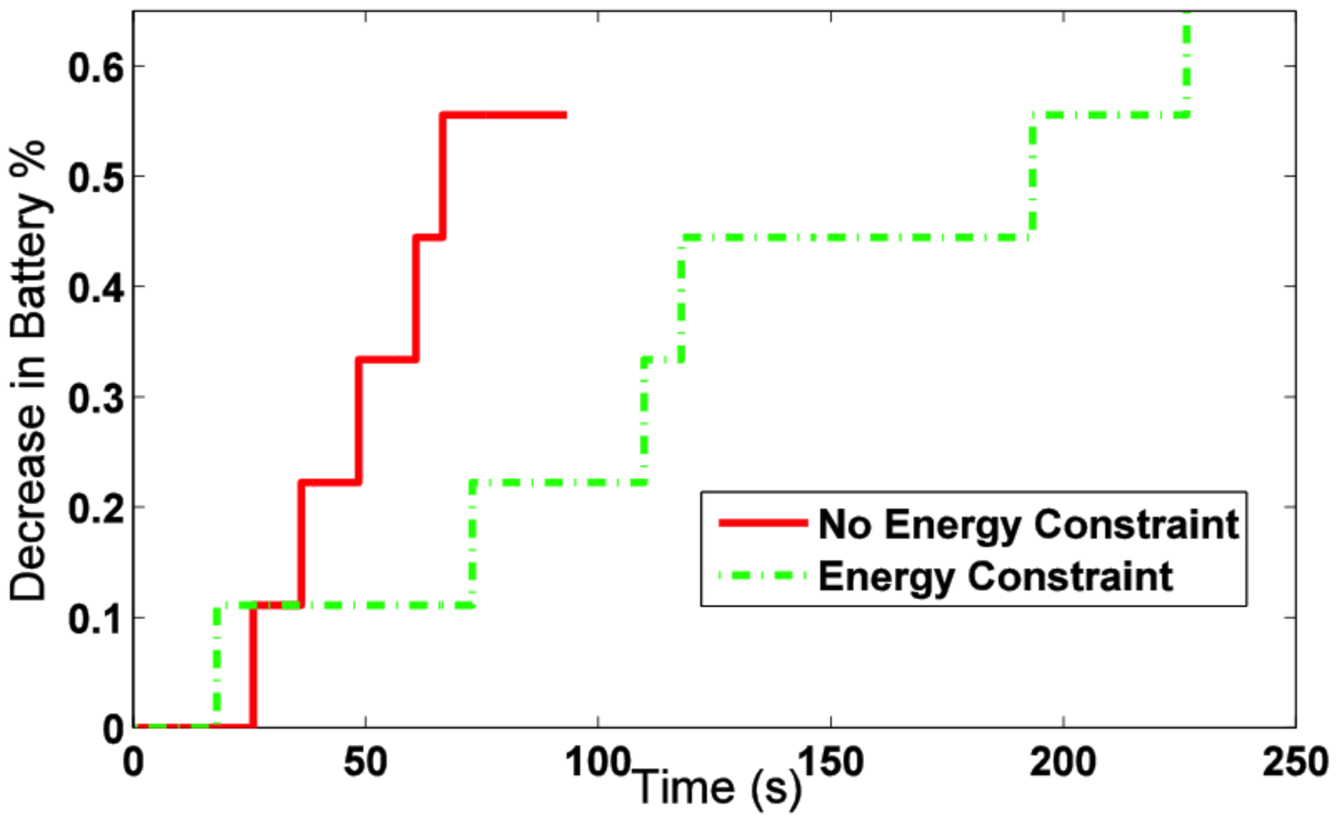}} } 
\subfigure[$\Oset(n)$]{ {\includegraphics[height = 32mm]{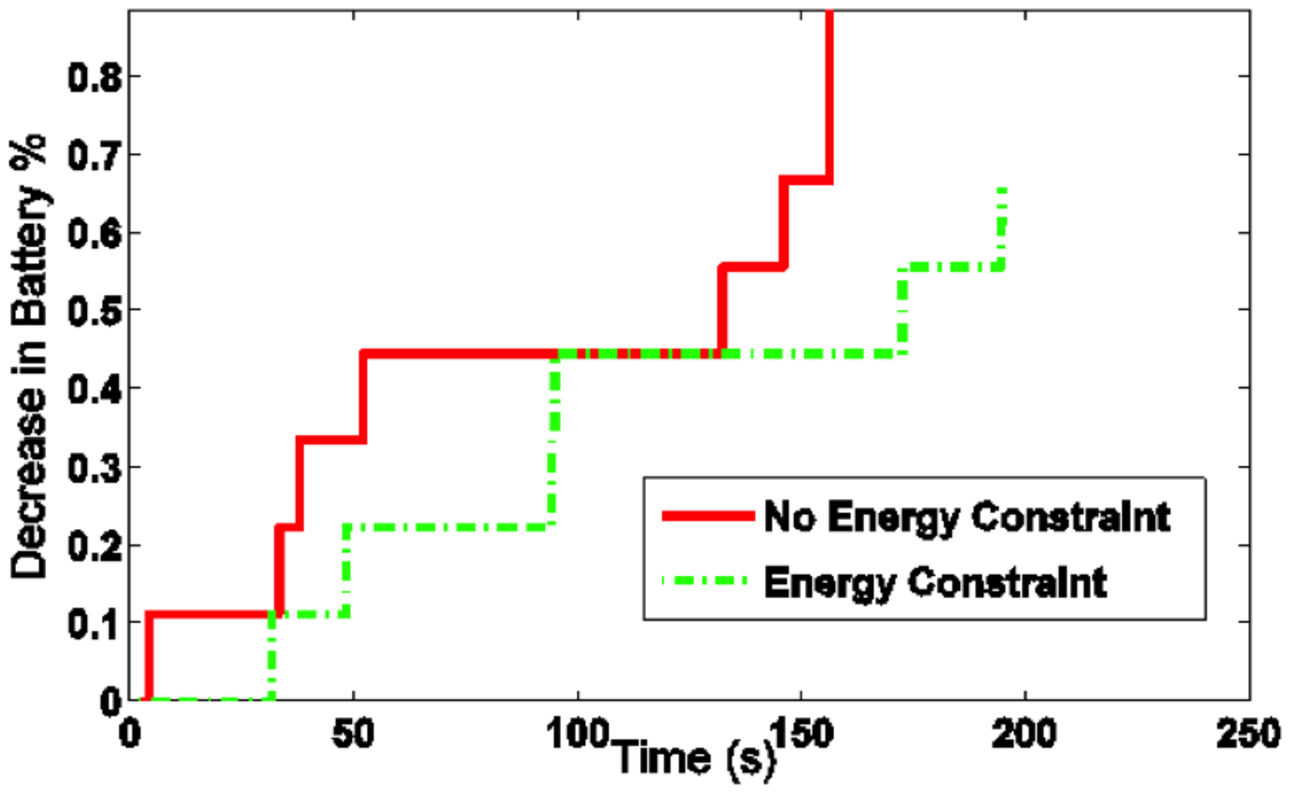}} } 
\subfigure[$\Oset(n^2)$ ]{ {\includegraphics[height = 32mm]{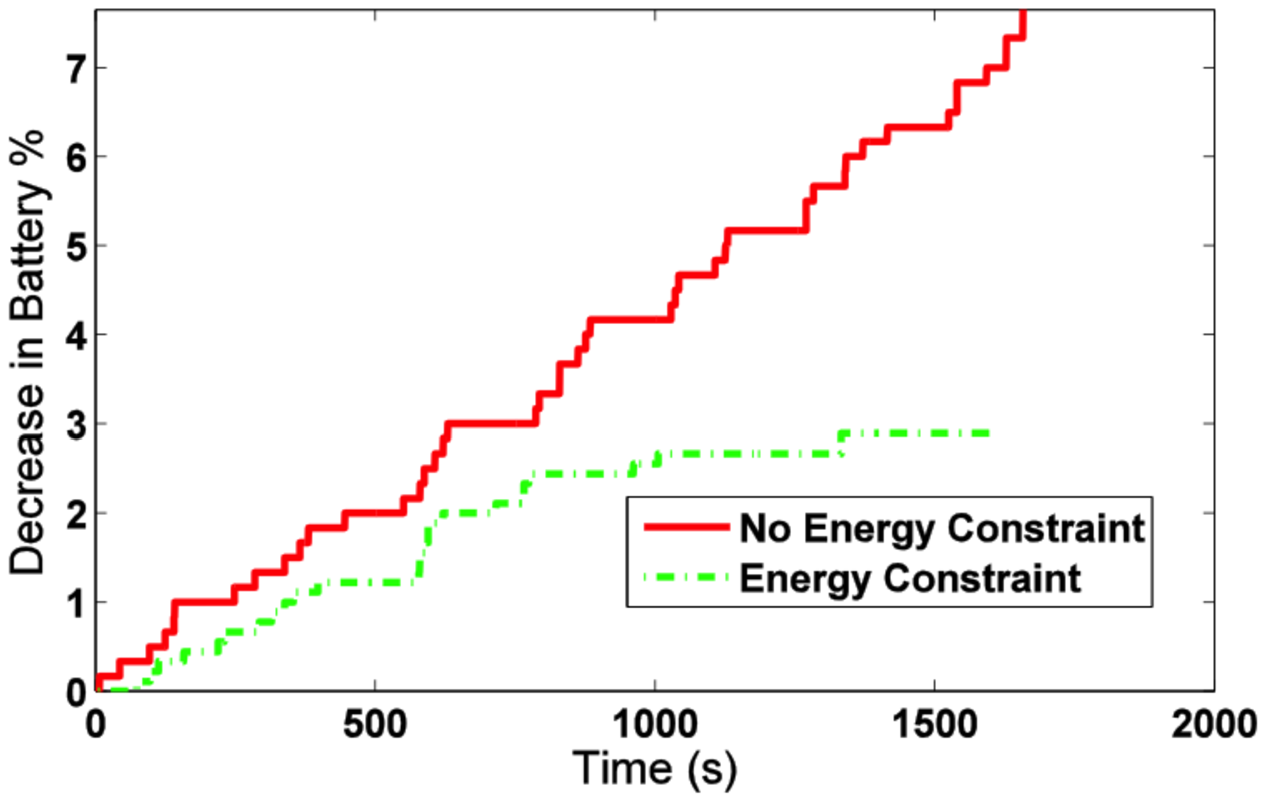}} } 
\vspace{-5pt}
\caption{Energy consumption of `Energy Constraint'' and ``No Energy Constraint'' scenarios over time for different computational complexities; $\Oset(1)$, $\Oset(n)$, and $\Oset(n^2)$. In this setup, the mobile devices are Android operating system (OS) \cite{AndroidDeveloper} based Nexus 7 tablets \cite{NexusTechSpecs}. The specific version of the Anroid OS is Android Lollipop 5.1.1. The devices have 16GB storage, 2GB RAM, Qualcomm Snapdragon S4 Pro, 1.5GHz CPU, and Adreno 320, 400MHz GPU. Packet size is $500B$. }
\vspace{-15pt}
\label{fig:energy_bottleneck}
\end{figure*}

In this section, we consider how energy can create a bottleneck. We consider the same setup as in Fig.~\ref{fig:complexityVsRate_new}(a), where data is transmitted from mobile device $D_1$ to another mobile device $D_2$. In this setup, $D_1$ transmits data after processing packets with computational complexities $\Oset(1)$, $\Oset(n)$, and $\Oset(n^2)$, where $n$ is the packet size, which is set to $500B$. We have two scenarios; ``Energy Constraint'' and ``No Energy Constraint''. In the energy constrained scenario, the assumption is that the battery level of $D_1$ is low, so it limits its transmission rate to $1Mbps$. On the other hand, in the unconstrained scenario, $D_1$ can transmit at the maximum possible transmission rates, which are reported in Fig.~\ref{fig:complexityVsRate_new}(b).

Fig.~\ref{fig:energy_bottleneck} shows energy consumption of the two scenarios over time for different computational complexities. As  seen, the energy consumption of the energy constrained version is much lower in all computational complexity levels, and the gap between the two scenarios increases when the computational complexity increases. From these results, we can conclude that devices can limit their transmission rates when their battery levels are limited. This will reduce energy consumption, but for the cost of limited transmission rate. Thus, energy levels of devices can become bottlenecks, even if bandwidth is not a bottleneck.

\section*{\label{sec:Th1_proof_unicast}Appendix B: Proof of Theorem~\ref{eec_theorem1}}
{\em Stability:}
Let $\boldsymbol H(t) = \{ \boldsymbol{{S}(t)}, \boldsymbol{{U}(t)}, \boldsymbol{Q(t)}, \boldsymbol{Z(t)} \}$, where $\boldsymbol{{S}(t)} = \{{S}_{n}(t)\}_{\forall n \in \Nset}$, $\boldsymbol{{U}(t)} = \{{U}_{n,k}(t)\}_{\forall n \in \Nset, k \in \Nset}$, $\boldsymbol{{Q}(t)} = \{{Q}_{n,k}(t)\}_{\forall n \in \Nset, k \in \Nset}$, and $\boldsymbol{{Z}(t)} = \{{Z}_{n,k}(t)\}_{\forall n \in \Nset, k \in \Nset}$.   

Let the Lyapunov function be;
\begin{align} \label{eq:appA_1}
& L(\boldsymbol H(t)) =  \sum_{n \in \Nset} S_{n}(t)^{2} + \sum_{n \in \Nset} \sum_{k \in \Nset} U_{n,k}(t)^{2} + \sum_{n \in \Nset} \sum_{k \in \Nset} \nonumber \\
& Q_{n,k}(t)^{2}  + \sum_{n \in \Nset} \sum_{k \in \Nset} Z_{n,k}(t)^{2} 
\end{align}
The Lyapunov drift is;
\begin{align} \label{eq:appA_2}
\Delta(\boldsymbol H(t)) = E[L(\boldsymbol H(t+1)) - L(\boldsymbol H(t)) | \boldsymbol H(t)]
\end{align} which is expressed as;
\begin{align} \label{eq:appA_3}
& \Delta(\boldsymbol H(t)) =  E[ \sum_{n \in \Nset} S_{n}(t+1)^{2} - \sum_{n \in \Nset} S_{n}(t)^{2} + \sum_{n \in \Nset} \sum_{k \in \Nset} \nonumber \\
& U_{n,k}(t+1)^{2} - \sum_{n \in \Nset} \sum_{k \in \Nset} U_{n,k}(t)^{2} + \sum_{n \in \Nset} \sum_{k \in \Nset} Q_{n,k}(t+1)^{2} \nonumber \\
\end{align}
\begin{align}
& - \sum_{n \in \Nset} \sum_{k \in \Nset} Q_{n,k}(t)^{2} + \sum_{n \in \Nset} \sum_{k \in \Nset} Z_{n,k}(t+1)^{2} - \sum_{n \in \Nset} \sum_{k \in \Nset} \nonumber \\
& Z_{n,k}(t)^{2} | \boldsymbol H(t)]
\end{align}
Considering the fact that $(\max[Q-b,0]+A)^{2} \leq Q^{2} + A^{2} + b^{2} + 2Q(A-b)$, (\ref{eq:appA_3}) is expressed as;
\begin{align}\label{eq:appA_3}
& \Delta(\boldsymbol H(t)) \leq  E\biggl[  \sum_{n \in \Nset} \biggl( S_{n}(t)^{2} + \Bigl(\sum_{k \in \Nset} x_{k,n}(t) \Bigr)^{2} + x_{n}(t)^{2} + \nonumber \\
& 2 S_n(t) \Bigl(x_n(t) - \sum_{k \in \Nset} x_{k,n}(t) \Bigr)  \biggr) - \sum_{n \in \Nset} S_{n}(t)^{2}  + \sum_{n \in \Nset} \sum_{k \in \Nset} \nonumber \\
&  \biggl( U_{n,k}(t)^{2} + x_{n,k}(t)^{2} + d_{n,k}(t)^{2} + 2 U_{n,k}(t) \Bigl(x_{n,k}(t) - \nonumber \\
& d_{n,k}(t) \Bigr)  \biggr) - \sum_{n \in \Nset} \sum_{k \in \Nset} U_{n,k}(t)^{2} 
+ \sum_{n \in \Nset} \sum_{k \in \Nset}  \biggl( Q_{n,k}(t)^{2} + \nonumber \\
& \Bigl(d_{n,k}(t)\alpha_{n,k}(t) \Bigr)^{2} + e_{n,k}(t)^{2} + 2 Q_{n,k}(t) \Bigl(d_{n,k}(t) \alpha_{n,k}(t) - \nonumber \\
&  e_{n,k}(t) \Bigr)  \biggr) - \sum_{n \in \Nset} \sum_{k \in \Nset} Q_{n,k}(t)^{2} + \sum_{n \in \Nset} \sum_{k \in \Nset}  \biggl( Z_{n,k}(t)^{2} + \nonumber \\
&  h_{n,k}(t)^{2} + e_{n,k}(t)^{2} + 2 Z_{n,k}(t) \Bigl(e_{n,k}(t)  -  h_{n,k}(t) \Bigr)  \biggr) - \nonumber \\
& \sum_{n \in \Nset} \sum_{k \in \Nset} Z_{n,k}(t)^{2}   | \boldsymbol H(t)\biggr]
\end{align} 
which is expressed as
\begin{align}\label{eq:appA_4}
& \Delta(\boldsymbol H(t)) \leq  E\biggl[  \sum_{n \in \Nset} \biggl( \Bigl(\sum_{k \in \Nset} x_{k,n}(t) \Bigr)^{2} + x_{n}(t)^{2} +  2 S_n(t) \Bigl( \nonumber \\
& x_n(t) - \sum_{k \in \Nset} x_{k,n}(t) \Bigr)  \biggr)  + \sum_{n \in \Nset} \sum_{k \in \Nset}   \biggl( x_{n,k}(t)^{2} + d_{n,k}(t)^{2} + \nonumber \\
& 2 U_{n,k}(t) \Bigl(x_{n,k}(t) -  d_{n,k}(t) \Bigr)  \biggr) + \sum_{n \in \Nset} \sum_{k \in \Nset}  \biggl(  \Bigl(d_{n,k}(t)\alpha_{n,k}(t) \Bigr)^{2} \nonumber \\
& + e_{n,k}(t)^{2} + 2 Q_{n,k}(t) \Bigl(d_{n,k}(t) \alpha_{n,k}(t) -  e_{n,k}(t) \Bigr)  \biggr) +  \sum_{n \in \Nset} \nonumber \\
& \sum_{k \in \Nset}  \biggl( h_{n,k}(t)^{2} + e_{n,k}(t)^{2} + 2 Z_{n,k}(t) \Bigl(e_{n,k}(t)  -  h_{n,k}(t) \Bigr)  \biggr) \nonumber \\
& | \boldsymbol H(t)\biggr]
\end{align} 

There always exists a finite positive constant $B$ satisfying
\begin{align}\label{eq:appA_5}
& B \geq E\biggl[  \sum_{n \in \Nset} \biggl( \Bigl(\sum_{k \in \Nset} x_{k,n}(t) \Bigr)^{2} + x_{n}(t)^{2}  \biggr)  + \sum_{n \in \Nset} \sum_{k \in \Nset}   \biggl( x_{n,k}(t)^{2} \nonumber \\
& + d_{n,k}(t)^{2}  \biggr) + \sum_{n \in \Nset} \sum_{k \in \Nset}  \biggl(  \Bigl(d_{n,k}(t)\alpha_{n,k}(t) \Bigr)^{2}  + e_{n,k}(t)^{2} \biggr) + \nonumber \\
& \sum_{n \in \Nset} \sum_{k \in \Nset}  \biggl( h_{n,k}(t)^{2} + e_{n,k}(t)^{2} \biggr)  | \boldsymbol H(t)\biggr]
\end{align}  because the maximum values of $x_n(t)$, $x_{n,k}(t)$, $d_{n,k}(t)$, $h_{n,k}(t)$, $e_{n,k}(t)$, and $\alpha_{n,k}(t)$ terms are bounded by finite positive constants by our EaCC algorithm. 

By taking into account (\ref{eq:appA_5}), (\ref{eq:appA_4}) is expressed as
\begin{align}\label{eq:appA_6}
& \Delta(\boldsymbol H(t)) \leq  B + E\biggl[  \sum_{n \in \Nset} \biggl( 2 S_n(t) \Bigl(  x_n(t) - \sum_{k \in \Nset} x_{k,n}(t) \Bigr)  \biggr) \nonumber \\
& + \sum_{n \in \Nset} \sum_{k \in \Nset}   \biggl(  2 U_{n,k}(t) \Bigl(x_{n,k}(t) -  d_{n,k}(t) \Bigr)  \biggr) + \sum_{n \in \Nset} \sum_{k \in \Nset} \nonumber \\
& \biggl(  2 Q_{n,k}(t) \Bigl(d_{n,k}(t) \alpha_{n,k}(t) -  e_{n,k}(t) \Bigr)  \biggr) +  \sum_{n \in \Nset}  \sum_{k \in \Nset}  \biggl( 2 Z_{n,k}(t) \nonumber \\
& \Bigl(e_{n,k}(t)  -  h_{n,k}(t) \Bigr)  \biggr) | \boldsymbol H(t)\biggr]
\end{align} 

The minimization of the right hand side of the drift inequality in (\ref{eq:appA_6}) corresponds to the decoder control in (\ref{eq:decoder}), energy control in (\ref{eq:energy}), and scheduling \& cooperation in (\ref{eq:schedulingCoop}).

If the arrival rates satisfy $E[x_{n}(t)] = A_{n}$ and $(A_{n})$ is inside the stability region $\Lambda$, then there exists a randomized policy with solution; $\oset{*}{x}_n(t)$, $\oset{*}{x}_{n,k}(t)$, $\oset{*}{d}_{n,k}(t)$, $\oset{*}{h}_{n,k}(t)$, and $\oset{*}{e}_{n,k}(t)$, satisfying
\begin{align}\label{eq:appA_7}
& -E[\sum_{k \in \Nset} \oset{*}{x}_{k,n}(t) - \oset{*}{x}_n(t)] \leq -\delta_1, \forall n \in \Nset \nonumber \\
& -E[\oset{*}{d}_{n,k}(t) - \oset{*}{x}_{n,k}(t)] \leq -\delta_2,  \forall n \in \Nset, k \in \Nset \nonumber \\
& -E[\oset{*}{e}_{n,k}(t) - \oset{*}{d}_{n,k}(t){\alpha}_{n,k}(t)] \leq -\delta_3,  \forall n \in \Nset, k \in \Nset \nonumber \\
& -E[\oset{*}{h}_{n,k}(t) - \oset{*}{e}_{n,k}(t)] \leq -\delta_4,  \forall n \in \Nset, k \in \Nset 
\end{align} where $\delta_1, \delta_2, \delta_3, \delta_4$ are positive small constants.

Since our EaCC algorithm minimizes the right hand side of (\ref{eq:appA_6}), the following inequalities satisfy: (i) $-E[\sum_{k \in \Nset} {x}_{k,n}(t) - {x}_n(t)] \leq -E[\sum_{k \in \Nset} \oset{*}{x}_{k,n}(t) - \oset{*}{x}_n(t)] \leq -\delta_1$, (ii) $-E[{d}_{n,k}(t) - {x}_{n,k}(t)] \leq -E[\oset{*}{d}_{n,k}(t) - \oset{*}{x}_{n,k}(t)] \leq -\delta_2$, (iii) $-E[{e}_{n,k}(t) - {d}_{n,k}(t){\alpha}_{n,k}(t)] \leq -E[\oset{*}{e}_{n,k}(t) - \oset{*}{d}_{n,k}(t){\alpha}_{n,k}(t)] \leq -\delta_3$, and (iv) $-E[{h}_{n,k}(t) - {e}_{n,k}(t)] \leq -E[\oset{*}{h}_{n,k}(t) - \oset{*}{e}_{n,k}(t)] \leq -\delta_4$. Thus, the following inequality satisfy
\begin{align}\label{eq:appA_8}
& \Delta(\boldsymbol H(t)) \leq  B - 2 \sum_{n \in \Nset} S_n(t) \delta_1 -  2 \sum_{n \in \Nset} \sum_{k \in \Nset} U_{n,k}(t) 
\delta_2 - \nonumber \\
& 2 \sum_{n \in \Nset} \sum_{k \in \Nset}  Q_{n,k}(t)  \delta_3 - 2 \sum_{n \in \Nset}  \sum_{k \in \Nset}   Z_{n,k}(t) \delta_4
\end{align}

Since there exists $\delta > 0$ satisfying $\delta \leq \min [\delta_1, \delta_2, \delta_3, \delta_4]$, the time average of the Lyapunov drift in (\ref{eq:appA_8}) is expressed as
\begin{align} \label{eq:appA_9}
& \limsup_{t \rightarrow \infty} \frac{1}{t} \sum_{\tau = 0}^{t-1} \frac{\Delta(\boldsymbol H(t))}{2} \leq \limsup_{t \rightarrow \infty} \frac{1}{t} \sum_{\tau=0}^{t-1} \biggl( \frac{B}{2} - \sum_{n \in \Nset} S_n(t) \delta - \nonumber \\
&  \sum_{n \in \Nset} \sum_{k \in \Nset} U_{n,k}(t)  \delta -  \sum_{n \in \Nset} \sum_{k \in \Nset}  Q_{n,k}(t)  \delta - \sum_{n \in \Nset}  \sum_{k \in \Nset}   Z_{n,k}(t) \delta \biggl)
\end{align} which leads to

\begin{align} \label{eq:appA_10}
& \limsup_{t \rightarrow \infty} \frac{1}{t} \sum_{\tau=0}^{t-1} \biggl(\sum_{n \in \Nset} S_n(t) +  \sum_{n \in \Nset} \sum_{k \in \Nset} \Bigl( U_{n,k}(t) +  Q_{n,k}(t)  + \nonumber \\
& Z_{n,k}(t) \Bigr)  \biggl) \leq \frac{B}{2\delta}
\end{align} concluding that the time average of the sum of the queues are bounded. This concludes the stability analysis part of the proof.

{\em Optimality:} 
Let us define a drift-plus-penalty function as $\Delta(\boldsymbol H(t)) - \sum_{k \in \Nset} M E[ g_{n}({x}_{n}(t)) | \boldsymbol H(t)]$ which is, considering the bound in (\ref{eq:appA_6}), expressed as
\begin{align}\label{eq:appA_11}
& \Delta(\boldsymbol H(t)) - \sum_{n \in \Nset} M E\biggl[ g_{n}({x}_{n}(t)) | \boldsymbol H(t) \biggr] \leq  B - 2E\biggl[  \sum_{n \in \Nset} \nonumber \\
& \biggl( S_n(t) \Bigl(   \sum_{k \in \Nset} x_{k,n}(t) - x_n(t) \Bigr)  \biggr)  + \sum_{n \in \Nset} \sum_{k \in \Nset}   \biggl(  U_{n,k}(t) \Bigl( d_{n,k}(t) \nonumber \\
& - x_{n,k}(t) \Bigr)  \biggr) + \sum_{n \in \Nset} \sum_{k \in \Nset} \biggl(  Q_{n,k}(t) \Bigl( e_{n,k}(t) - d_{n,k}(t) \alpha_{n,k}(t) \Bigr)  \biggr) \nonumber \\
& +  \sum_{n \in \Nset}  \sum_{k \in \Nset}  \biggl( Z_{n,k}(t)  \Bigl( h_{n,k}(t) - e_{n,k}(t)  \Bigr)  \biggr) | \boldsymbol H(t)\biggr] - \sum_{n \in \Nset} M \nonumber \\
& E\biggl[ g_{n}({x}_{n}(t)) | \boldsymbol H(t) \biggr] 
\end{align} 

Note that the minimization of the right hand side of the drift inequality in (\ref{eq:appA_11}) corresponds the flow control part of EaCC in Section~\ref{sec:CoopComp} as well as the decoder control in (\ref{eq:decoder}) , energy control in (\ref{eq:energy}), and scheduling \& cooperation in (\ref{eq:schedulingCoop}). Since there exists a randomized policy as discussed in the stability part above, the right hand side of (\ref{eq:appA_11}) is bounded as

\begin{align}\label{eq:appA_11}
& \Delta(\boldsymbol H(t)) - \sum_{n \in \Nset} M E\biggl[ g_{n}({x}_{n}(t)) | \boldsymbol H(t) \biggr] \leq  B - 2 \sum_{n \in \Nset} S_n(t) \nonumber \\
& \delta  -  2 \sum_{n \in \Nset} \sum_{k \in \Nset} \Bigl( U_{n,k}(t) +  Q_{n,k}(t) +  Z_{n,k}(t) \Bigr) \delta  - \sum_{n \in \Nset} M \nonumber \\
& E[g_{n}(A_{n}+\delta)]
\end{align} where $\sum_{n \in \Nset} g_{n}(A_{n})$ is the maximum time average of the sum utility function that can be achieved by any control policy that stabilizes the system. We can rewrite (\ref{eq:appA_11}) as 
\begin{align} \label{eq:appA_12}
& \limsup_{t \rightarrow \infty} \frac{1}{t} \sum_{\tau = 0}^{t-1} \biggl[\Delta(\boldsymbol H(\tau))  - \sum_{n \in \Nset} M E[g_{n}(x_{n}(\tau))] \biggr] \leq  \limsup_{t \rightarrow \infty} \nonumber \\
& \frac{1}{t} \sum_{\tau = 0}^{t-1} \biggl[ B - 2 \sum_{n \in \Nset} S_n(t)  \delta  -  2 \sum_{n \in \Nset} \sum_{k \in \Nset} \Bigl( U_{n,k}(t) +  Q_{n,k}(t) + \nonumber \\
&  Z_{n,k}(t) \Bigr) \delta  - \sum_{n \in  \Nset} M U_{n}(A_{n}  +\delta) \biggr]
\end{align}

Let us first consider the stability of the queues. If both sides of (\ref{eq:appA_12}) are divided by $\delta$ and the terms are arranged, we have
\begin{align}\label{eq:appA_13}
& \limsup_{t \rightarrow \infty} \frac{1}{t} \sum_{\tau=0}^{t-1} \biggl(\sum_{n \in \Nset} S_n(t) +  \sum_{n \in \Nset} \sum_{k \in \Nset} \Bigl( U_{n,k}(t) +  Q_{n,k}(t)  + \nonumber \\
& Z_{n,k}(t) \Bigr)  \biggl) \leq \frac{B}{2\delta} +  \limsup_{t \rightarrow \infty} \frac{1}{t} \sum_{\tau = 0}^{t-1} \biggl[ \sum_{n \in \Nset}  \frac{M}{\delta} E[g_{n}(x_{n}(\tau))]  \biggr] - \nonumber \\
& \sum_{n \in \Nset} \frac{Mg_{n}(A_{n}+\delta)}{\delta}. 
\end{align} This concludes that EaCC stabilizes the queues in the system when the flow control algorithm in Section~\ref{sec:CoopComp} is employed. Next, we consider the optimality of EaCC. If both sides of (\ref{eq:appA_12}) are divided by $M$, we have
\begin{align} \label{eq:appA_14}
& \limsup_{t \rightarrow \infty} \frac{1}{t} \sum_{\tau = 0}^{t-1} \biggl[ - \sum_{n \in \Nset} E[g_{n} (x_{n} (\tau)) ] \biggr] \leq
\limsup_{t \rightarrow \infty}  \frac{1}{t} \sum_{\tau = 0}^{t-1} \biggl[ \frac{B}{M} - \nonumber \\
& 2 \sum_{n \in \Nset} S_n(t)  \frac{\delta}{M}  -  2 \sum_{n \in \Nset} \sum_{k \in \Nset} \Bigl( U_{n,k}(t) +  Q_{n,k}(t) +  Z_{n,k}(t) \Bigr) \nonumber \\
& \frac{\delta}{M}  - \sum_{n \in  \Nset} g_{n}(A_{n}  +\delta) \biggr] 
\end{align} which is expressed as
\begin{align} \label{eq:appA_15}
& \limsup_{t \rightarrow \infty} \frac{1}{t} \sum_{\tau = 0}^{t-1} \biggl[ \sum_{n \in \Nset} E[g_{n} (x_{n} (\tau)) ] \biggr] \geq
\limsup_{t \rightarrow \infty}  \frac{1}{t} \sum_{\tau = 0}^{t-1} \biggl[ \sum_{n \in  \Nset} \nonumber \\
& g_{n}(A_{n}  +\delta)  - \frac{B}{M} +  2 \sum_{n \in \Nset} S_n(t)  \frac{\delta}{M}  +  2 \sum_{n \in \Nset} \sum_{k \in \Nset} \Bigl( U_{n,k}(t) \nonumber \\
& +  Q_{n,k}(t) +  Z_{n,k}(t) \Bigr)  \frac{\delta}{M}  \biggr] 
\end{align} Since $\limsup_{t \rightarrow \infty}  \frac{1}{t} \sum_{\tau = 0}^{t-1} \biggl[ \sum_{n \in  \Nset}  2 \sum_{n \in \Nset} S_n(t)  \frac{\delta}{M}  +  2 \sum_{n \in \Nset} \sum_{k \in \Nset} \Bigl( U_{n,k}(t) +  Q_{n,k}(t) +  Z_{n,k}(t) \Bigr)  \frac{\delta}{M}  \biggr] \geq 0$, the following inequality holds
\begin{align} \label{eq:appA_16}
& \limsup_{t \rightarrow \infty} \frac{1}{t} \sum_{\tau = 0}^{t-1} \biggl[ \sum_{n \in \Nset} E[g_{n} (x_{n} (\tau)) ] \biggr] \geq
 \sum_{n \in  \Nset}  g_{n}(A_{n}  +\delta)  - \frac{B}{M}. 
\end{align} This proves that the flow rates achieved by EaCC converge to the utility optimal operating point with increasing $M$. This concludes the optimality part of the proof.

\end{document}